\documentclass{aa}  

\usepackage[version=4]{mhchem}
\usepackage{graphicx}
\usepackage{txfonts}
\usepackage{multirow}
\usepackage{xcolor}
\usepackage{ulem}
\usepackage[colorlinks=true,linkcolor=black,citecolor=blue,urlcolor=blue,draft=false]{hyperref}

\begin{document} 

   \title{Disentangling the dust and gas contributions of the JWST/MIRI spectrum of Sz\,28}

   \author{T. Kaeufer\inst{1}\fnmsep\inst{2}\fnmsep\inst{3}\fnmsep\inst{4},
           P. Woitke\inst{1},
           I. Kamp\inst{2}, 
           J. Kanwar\inst{2}\fnmsep\inst{1}\fnmsep\inst{4}, \and  
           M. Min\inst{3}}
   \institute{
   Space Research Institute, Austrian Academy of Sciences, Schmiedlstrasse 6, 8042 Graz, Austria \\ \email{till.kaeufer@oeaw.ac.at}
        \and
    Kapteyn Astronomical Institute, University of Groningen, PO Box 800, 9700 AV Groningen, The Netherlands
        \and
   SRON Netherlands Institute for Space Research, Niels Bohrweg 4, 2333CA Leiden, The Netherlands
        \and
        Institute for Theoretical Physics and Computational Physics, Graz University of Technology, Petersgasse 16, 8010 Graz, Austria
        }

   \date{Received 28 May 2024 / Accepted 8 August 2024}

 
  \abstract
   {
   Recent spectra of protoplanetary disks around very low-mass stars (VLMS), captured by the Mid-InfraRed Instrument (MIRI) on board the \textit{James Webb} Space Telescope (JWST), reveal a rich carbon chemistry. Current interpretations of these spectra are based on 0D slab models and provide valuable estimates for molecular emission temperatures and column densities in the innermost disk ($\rm radius \lesssim 1\,\rm au$). However, the established fitting procedures and simplified models are challenged by the many overlapping gas features.}
   {We aim to simultaneously determine the molecular and the dust composition of the disk around the VLMS Sz\,28 in a Bayesian way.}
   {We model the JWST/MIRI spectrum of Sz\,28 up to $17\,\rm \mu m$ using the Dust Continuum Kit with Line emission from Gas (DuCKLinG). Systematically excluding different molecules from the Bayesian analysis allows for an evidence determination of all investigated molecules and isotopologues. We continue by examining the emission conditions and locations of all molecules, analysing the differences to previous 0D slab fitting, and analysing the dust composition.}
   {We find very strong Bayesian evidence for the presence of \ce{C2H2}, \ce{HCN}, \ce{C6H6}, \ce{CO2}, \ce{HC3N}, \ce{C2H6}, \ce{C3H4}, \ce{C4H2}, and \ce{CH4} in the JWST/MIRI spectrum of Sz\,28. Additionally, we identify \ce{CH3} and find tentative indications for \ce{NH3}. There is no evidence for water in the spectrum. However, we show that column densities of up to $2\times10^{17}\,\rm cm^{-2}$ could be hidden in the observational noise if assuming similar emission conditions of water as the detected hydrocarbons. Contrary to previous 0D slab results, a \ce{C4H2} quasi-continuum is robustly identified. We confirm previous conclusions that the dust in Sz\,28 is highly evolved, with large grains ($5\,\rm \mu m$) and a high crystallinity fraction being retrieved. We expect some of the stated differences to previous 0D slab fitting results to arise from an updated data reduction of the spectrum, but also due to the different modelling process. The latter reason underpins the need for more advanced models and fitting procedures.}
   {}


   \authorrunning{T.~Kaeufer et al.}
   
   \maketitle
%
\section{Introduction\label{sec:intro}}

Very low-mass stars (VLMS) \citep{Liebert1987} are known to host many terrestrial exoplanets \citep[e.g.][]{Sabotta2021,Ment2023}. With the formation conditions strongly influencing the planetary compositions \citep{Oberg2011,Molliere2022}, the planet-forming regions around these stars can provide crucial information to improve our understanding of planet formation.
However, most studies analysing the molecular composition with the \textit{Spitzer} space telescope focused on T\,Tauri and Herbig Ae/Be disks that provide stronger signals \citep{Salyk2011,Carr2011}. 

Pioneering work regarding the molecular content of the disks around VLMS established an underabundance of \ce{HCN} compared to \ce{C2H2} \citep{Pascucci2009} and a high C/O ratio in the inner disk compared to T\,Tauri disks \citep{Pascucci2013}. This illustrates that these objects are not just miniature version of their larger T\,Tauri and Herbig counterparts. 

The improved sensitivity and spectral resolution of the medium resolution spectrograph of the Mid-InfraRed Instrument (MIRI) on board the \textit{James Webb} Space Telescope (JWST) compared to \textit{Spitzer} allows for a detailed look at the dust and gas properties in the inner disks around VLMS. Analyses based on 0D slab models detected a large collection of hydrocarbons \citep{Tabone2023,Arabhavi2024,Kanwar2024}. Except for weak \ce{H2O} features in J1605321-1933159 \citep[hereafter J160532,][]{Tabone2023} no water is detected in the inner disks around any of these VLMS. 
This confirms the previously determined high C/O ratio in the inner disks around VLMS \citep{Pascucci2013}. 

JWST/MIRI spectra, especially but not only for VLMS, show many overlapping molecular features. Describing these spectra with a consistent thermo-chemical disk structure is connected to substantial computational challenges \citep{Woitke2024}. Therefore, 0D slab models that determine the local thermal equilibrium (LTE) flux of molecules individually for given column densities, temperatures, and emitting areas are a valuable fast tool for a first characterization of the molecular emission conditions \citep[e.g.][]{Grant2023,Schwarz2024,Munoz2024}.

While single attempts have been made to compare slab models to JWST/MIRI spectra in a Bayesian way \citep{Munoz2024}, most slab conditions have been selected using a sequence of $\chi^2$ minimizations to fit molecules or groups of molecules, each time subtracting the resulting fit from the observed spectrum \citep[e.g.][]{Gasman2023,Xie2023}. This becomes especially challenging for VLMS that exhibit many overlapping molecular features.

In this paper, we analyse the dust and gas composition of Sz\,28, the only known source to date with many hydrocarbons and dust emission features in the JWST/MIRI spectrum, in a Bayesian way. 

Sz\,28 is a M5.5 star with a mass of $0.12\,\rm M_\odot$, a temperature of about $3060\,\rm K$ \citep{Manara2017}, and a luminosity of $0.04\,\rm L_\odot$ \citep{Kanwar2024}. This VLMS at a distance of $192.2\,\rm pc$ \citep{Gaia2020} is part of the Chameleon I region.
A \textit{Spitzer} analysis revealed a flat-topped silicate feature at $10\,\rm \mu m$ and crystalline features hinting to more processed dust compared to the ISM \citep{Pascucci2009}. Additionally, a clear \ce{C2H2} feature has been identified that is consistent with the trend of cool stars having stronger \ce{C2H2} emission compared to \ce{HCN} \citep{Pascucci2009}.

We compare the Dust continuum Kit with Line emission from Gas \citep[DuCKLinG,][]{Kaeufer2024} models to the JWST/MIRI spectrum of Sz\,28. These 1D models, optimised for speed, describe the dust and gas emission at the same time using temperature gradients for both components and additional column density gradients for the gaseous emission. We quantify with our Bayesian approach the evidence for molecules and isotopologues present in the spectrum. The simultaneous description of dust and gas helps us to disentangle the dust and gas contributions, including gas quasi-continua which are easily mistaken for dust continua. Additionally, the 1D model provides some limited information about the location in the disk associated with the emission of certain molecules. We accompany these findings by an analysis of the water abundance in Sz\,28 and the dust composition.

This paper is structured in the following way. In Sect.~\ref{sec:method}, we describe the model (Sect.~\ref{sec:duckling}), observation (Sect.~\ref{sec:obs}), and Bayesian strategy (Sect.~\ref{sec:fitting}). The results are presented in Sect.~\ref{sec:results}, divided into the evidence for different molecules (Sect.~\ref{sec:mol_select}) and istopologues (Sect.~\ref{sec:iso}) followed by an analysis of the emitting conditions of the emitting species (Sect.~\ref{sec:final_fit}). Thereafter, we focus on the emission locations in the disk (Sect.~\ref{sec:location}), constrain the water abundance (Sec.~\ref{sec:water_limit}), analyse the robustness of the detected \ce{C4H2} quasi-continuum (Sect.~\ref{sec:quasi-continuum}), and examine the dust composition (Sect.~\ref{sec:dust_composition}), before concluding the paper with a summary of the main finding in Sect.~\ref{sec:summary}.

\section{Method\label{sec:method}}

In this section, we describe the process to determine the gas and dust emitting properties in a protoplanetary disk based on a JWST/MIRI spectrum. The observation is compared to the flux of DuCKLinG \citep{Kaeufer2024} models using a Bayesian sampling algorithm.

\subsection{DuCKLinG\label{sec:duckling}}

The DuCKLinG model describes the dust and gas emission of a protoplanetary disk as the sum of various 1D components. Next to a star, the dust is represented by three components (rim, midplane, and surface layer) with one further component describing the molecular emissions.

For a detailed description of the components, see \cite{Kaeufer2024}. In this section, we shortly summarise the components and highlight the changes made compared to \cite{Kaeufer2024}.

The stellar flux is given by a stellar PHOENIX spectrum \citep{Brott2005} using the stellar parameters as introduced for Sz\,28 by \cite{Kanwar2024}.

The simple black body with a temperature $T_{\rm rim}$ represents the inner rim of the dust disk.
The rest of the optically thick dust emission is described by the midplane component, integrating the back body emission over a radial temperature power law (exponent $q_{\rm mid}$) from $T_{\rm max}^{\rm mid}$ at the inner edge to $T_{\rm min}^{\rm mid}$ at the outer one.

The dust surface layer component represents the optically thin dust emission. It is described by a superposition of dust species with individual opacities and column densities, and a common radial power law (from $T_{\rm max}^{\rm sur}$ to $T_{\rm min}^{\rm sur}$ with an exponent $q_{\rm sur}$) for the dust emission temperature.
The exact spectral profile of emission from small dust particles is very sensitive to the particle shape. Especially, since here we aim to capture small details in the gas emission on top of the dust continuum, it is crucial to model the dust emission as accurately as possible. Computationally fast approximations capture the essence, but not always the details of emissivities measured in the laboratory \citep[see e.g.][]{Mutschke2009}. Here we use the porous Gaussian Random Field particle shape model \citep{Min2007, Grynko2003, Shkuratov2005}. We compute the optical properties of the particles applying the Discrete Dipole Approximation (DDA) using the code \texttt{DDSCAT} \citep{Draine2013}. Note that DDA is the only computational method that allows taking into account the anisotropy of the refractive index of the crystalline silicates, allowing for a more detailed treatment of the exact positions and shapes of the resonances created in the crystal structure of these minerals. The references for the laboratory measurements for the refractive indices used are: amorphous silicate with Mg$_2$SiO$_4$ composition \citep[am Mg-olivine,][]{Henning1996},
amorphous silicate with MgSiO$_3$ composition \citep[am Mg-pyroxene,][]{Dorschner1995},
amorphous SiO$_2$ \citep[silica,][]{Henning1997},
crystalline forsterite \citep{Servoin1973}, and
crystalline enstatite \citep{Jager1998}.

The dust components are based on work by \cite{Juhasz2009,Juhasz2010}. \cite{Kaeufer2024} expanded this model with a gaseous component that introduces a radial column density and temperature power laws for every molecule (from $T_{\rm max}^{\rm mol}$/$\Sigma_{\rm tmax}^{\rm mol}$ to $T_{\rm min}^{\rm mol}$/$\Sigma_{\rm tmin}^{\rm mol}$). While the radial temperature slope ($q_{\rm emis}$) is set globally for all species, the molecules emit within individual radial/temperature ranges. Similarly, the column density power law is defined individually per molecule.

The underlying specific intensity that determine the flux per molecule are estimated by linear interpolation in a grid of 0D LTE slab models introduced by \cite{Arabhavi2024}. The models in this grid are spaced in temperature and column density steps of $25\,\mathrm{K}$ and $1/6\,\rm dex$, respectively. The grid extends from column densities of $10^{14}\rm cm^{-2}$ to $10^{24.5}\rm cm^{-2}$. For all molecules, the grid starts at temperatures of $25\,\mathrm{K}$ and continues up to $1500\,\mathrm{K}$ with the exceptions of \ce{C3H4} and \ce{C6H6} that extend to $600\,\mathrm{K}$ only. For a detailed description, see \cite{Arabhavi2024}. 
In this study, the slab output is convolved to the MIRI resolution varying with wavelength ($R=3500,3000,2500,1500$ for channels 1 to 4, respectively).

The total model flux sums up the flux from the stellar, dust, and molecular components. The independent treatment of every component means that no interaction between them is taken into account. This means that the model cannot describe optical depth effects of optically thick gas emission on the total flux and absorption features of gas. Similarly, the dust opacities do not influence the flux from the gaseous component which optical depths are calculated for every molecule individually using slab models. However, these limitations come with the benefit of high computational speed.

\subsection{Observational data\label{sec:obs}}

\begin{figure}[t]
    \centering
    \includegraphics[width=\linewidth]{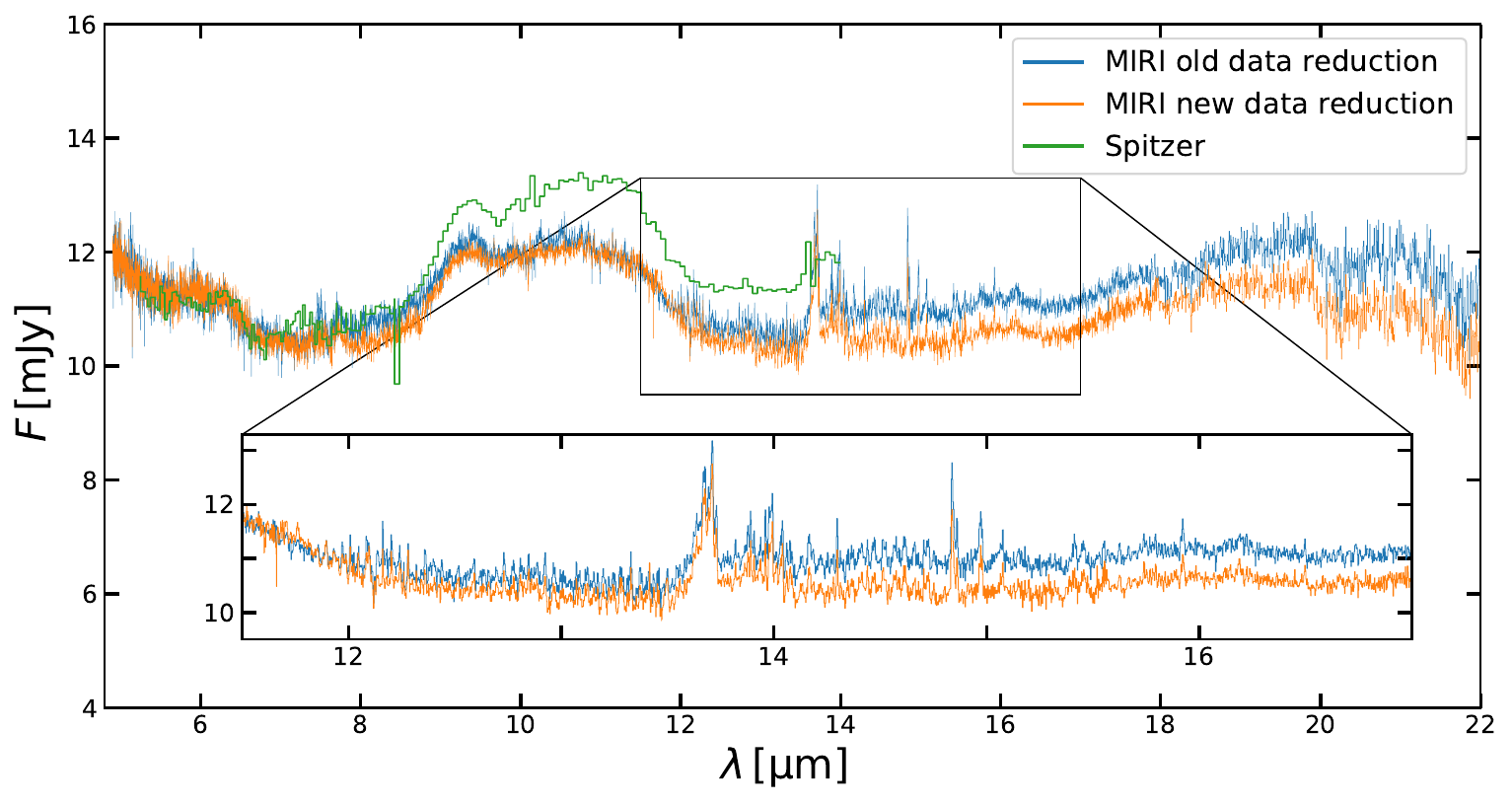}
    \caption{The MIRI spectrum of Sz\,28 newly reduced (JWST standard pipeline version 1.13.4, orange) and as shown by \cite{Kanwar2024} (version 1.11.1, blue). The \textit{Spitzer} spectrum of the same source is shown in green.}
    \label{fig:spec_datareduction}
\end{figure}

The JWST/MIRI spectrum of Sz\,28 reduced with version 1.11.1 of the JWST standard pipeline has been published by \cite{Kanwar2024} who estimated the lower limit of the observational uncertainty with the Exposure Time Calculator to be $0.1\,\rm mJy$. Since then several improvements to the data reduction have been made. Therefore, we show an updated spectrum \citep[reduced with version 1.13.4,][]{Bushouse2023} in this study (Fig.~\ref{fig:spec_datareduction}). While a detailed description of the data reductions is given by \cite{Kanwar2024}, we list here the improvements made. The JWST standard pipeline uses CRDS context of jwst\_1215.pmap with the VIP package version 1.4.0 \citep{GomezGonzalez2017, Christiaens2023}. 
Among the improvements of the updated pipeline are a correction of the spectral leak (at about $12.2\,\rm \mu m$) and an additional bad pixel correction leading to a higher signal-to-noise ratio.  
Fig.~\ref{fig:spec_datareduction} illustrates the improvements of the new data reduction compared to the one used by \cite{Kanwar2024}. The overall flux levels differ due to the new calibration files. Similarly, the line widths of weak features change slightly. All of this results in changes in the retrieved molecular conditions.

\subsection{Fitting procedure\label{sec:fitting}}

We use the UltraNest package \citep{Buchner2021} which utilizes the Monte Carlo algorithm MLFriends \cite{Buchner2016,Buchner2019} to derive the posterior distributions of the model parameters. This algorithm is optimised for robustness, especially for highly dimensional parameter spaces. The details are provided in the references above.

A classical Gaussian likelihood function $\mathcal{L}$ is used to evaluate the differences between the model spectrum $F_{\rm i,\rm model}$ and the observations $F_{\rm i,\rm obs}$ with uncertainties $\sigma_{\rm i,\rm obs}$. The index  $i\!=\!1\,...\,N_{\rm obs}$ enumerates all wavelength points in the selected wavelength range:

\begin{align}
    \mathcal{L} = \prod_{i=1}^{N_{\rm obs}}
    \frac{1}{\sqrt{2\pi\, \sigma_{i,\rm obs}^2}}
    \exp{\left(-\frac{\left(F_{i,\rm model} 
    -F_{i,\rm obs}\right)^2}
    {2\,\sigma_{i,\rm obs}^2}\right)} \ .
    \label{eq:likelihood_classic}
\end{align}

We derive the observational uncertainty as a free (non-linear) parameter. This avoids a manual uncertainty determination based on line-free wavelength regions of the observation. The parameter $a_{\rm obs}$ defines $\sigma_{\rm i,obs}$ as proportional to the flux of the observation, 
\begin{align}
    \sigma_{i,\rm obs}= a_{\rm obs}\, F_{i,\rm obs}. \label{eq:flux_uncertainty}
\end{align}
The total flux of the model is rebinned to the observation's wavelength points ($F_{\rm i,\rm model}$) using SpectRes Python package \citep{Carnall2017}.

We adopt the approach described in \cite{Kaeufer2024} to reduce the complexity of the parameter space and, consequently, the computational cost of the retrieval process. This technique employs a non-negative least square (NNLS) solver \citep{scipy2020} to determine the linear parameters namely the dust abundances, the emitting areas of each molecule, the rim, and midplane. UltraNest creates a set of non-linear parameters, while the remaining linear parameters are determined through NNLS. The total set of parameters is then used to calculate the likelihood of the model with respect to the observation. Therefore, the linear parameters are not determined in a fully Bayesian way, but their posterior is approximated. However, \cite{Kaeufer2024} shows that this significant improvement in computational speed (a factor of $80$ for a mock retrieval) leaves the posterior of the molecular parameters basically unaffected, with only the uncertainties of the retrieved dust composition decreasing slightly.

\section{Results\label{sec:results}}

We aim to determine the molecular and the dust properties of Sz\,28 based on a large wavelength region (from $4.9 \,\rm \mu m$ to $17\,\rm \mu m$) of the JWST/MIRI spectrum. For that, we need to determine the molecules relevant to describe the spectrum of Sz\,28 in addition to the posterior of the retrieved parameters.
We adopt an interactive process to achieve this goal. In a first Bayesian analysis we include a large number of molecules. This includes molecules already detected in Sz\,28 \citep[\ce{C2H2}, \ce{HCN}, \ce{C6H6}, \ce{CO2}, \ce{HC3N}, \ce{C2H6}, \ce{C3H4}, \ce{C4H2}, \ce{CH3}, and \ce{CH4},][]{Kanwar2024}, but also potential molecules which existence or non-existence we want to evaluate (\ce{H2O}, \ce{CO}, \ce{NH3}, \ce{C2H4}). After that, we exclude each molecule individually and evaluate the evidence for their presence using the Bayes factor (Sect.~\ref{sec:mol_select}). Then, this process is repeated for the isotopologues for which we have spectroscopic data (Sect.~\ref{sec:iso}). This selection of molecules and isotopologues results in a final fit, that only includes the detectable species, for which the molecular emission conditions are described in Sect.~\ref{sec:final_fit}.

\subsection{Molecule selection\label{sec:mol_select}}

The prior ranges of all free parameters that are fitted with UltraNest are listed in Table~\ref{tab:prior_full_fit}. The dust temperature prior ranges are chosen to avoid convergence towards the prior's edges. All priors of the power law exponents are limited between $-1$ and $-0.1$, to include known literature values ($-0.6$ and $-0.4$, and values from $-0.95$ to $-0.5$ from \cite{Brittain2023} and \cite{Fedele2016}, respectively).
The prior ranges of all molecular parameters are determined by the grid's extent of the underlying slab models from \cite{Arabhavi2024}.

\begin{table}[t]
    \caption{Prior distributions of free parameters to fit Sz\,28 with the full complexity model.}
    \label{tab:prior_full_fit}
    \centering
    \begin{tabular}{l|l|l|l}
\hline \hline & & & \\[-1.9ex] 
Parameter & Prior & Parameter & Prior \\ \hline  
  & & & \\[-1.9ex] 
$T_{\rm rim}\,\rm [K]$ & $\mathcal{U}(800,1600)$ &$\Sigma_{\rm tmin}^{\rm mol}\tablefootmark{(1)}\,\rm [cm^{-2}]$ & $\mathcal{J}(10^{14},10^{24})$ \\ 
  & & & \\[-1.9ex] 
$T^{\rm sur}_{\rm min}\,\rm [K]$ & $\mathcal{U}(10,1000)$ &$\Sigma_{\rm tmax}^{\rm mol}\tablefootmark{(1)}\,\rm [cm^{-2}]$ & $\mathcal{J}(10^{14},10^{24})$ \\ 
  & & & \\[-1.9ex] 
$T^{\rm sur}_{\rm max}\,\rm [K]$ & $\mathcal{U}(50,1600)$ &$T_{\rm max}^{\rm mol}\tablefootmark{(1)}\,\rm [K]$ & $\mathcal{U}(25,1500)$ \\  
 & & & \\[-1.9ex] 
$T^{\rm mid}_{\rm min}\,\rm [K]$ & $\mathcal{U}(10,1000)$ &$T_{\rm min}^{\rm mol}\tablefootmark{(1)}\,\rm [K]$ & $\mathcal{U}(25,1500)$ \\  
 & & & \\[-1.9ex] 
$T^{\rm mid}_{\rm max}\,\rm [K]$ & $\mathcal{U}(300,1600)$ &$\Sigma_{\rm tmin}^{\rm mol}\tablefootmark{(2)}\,\rm [cm^{-2}]$ & $\mathcal{J}(10^{14},10^{24})$ \\ 
  & & & \\[-1.9ex] 
$q_{\rm mid}$ & $\mathcal{U}(-1,-0.1)$ &$\Sigma_{\rm tmax}^{\rm mol}\tablefootmark{(2)}\,\rm [cm^{-2}]$ & $\mathcal{J}(10^{14},10^{24})$ \\ 
  & & & \\[-1.9ex] 
$q_{\rm sur}$ & $\mathcal{U}(-1,-0.1)$ &$T_{\rm max}^{\rm mol}\tablefootmark{(2)}\,\rm [K]$ & $\mathcal{U}(25,600)$ \\  
 & & & \\[-1.9ex] 
$q_{\rm emis}$ & $\mathcal{U}(-1,-0.1)$ &$T_{\rm min}^{\rm mol}\tablefootmark{(2)} \,\rm [K]$& $\mathcal{U}(25,600)$ \\  
 & & & \\[-1.9ex] 
$a_{\rm obs}$ & $\mathcal{J}(10^{-5},10^{-1})$ & & \\
    \end{tabular}
        \tablefoot{\\
        $\mathcal{U}(x,y)$ and $\mathcal{J}(x,y)$ denote uniform and log-uniform priors in the range from $x$ to $y$, respectively.\\
\tablefoottext{1}{Same prior used for \ce{H2O}, \ce{NH_3}, \ce{C_2H_4}, \ce{CH_4}, \ce{C_2H_6}, \ce{HCN}, \ce{C_4H_2}, \ce{C_2H_2}, \ce{CO2}, \ce{HC_3N}, \ce{CO}.}\\
\tablefoottext{2}{Same prior used for \ce{CH_3}, \ce{C_3H_4}, \ce{C_6H_6}.}

}
\end{table}

The distance to the object is fixed to $192.2\,\rm pc$ \citep{Gaia2020}. We do not account for the inclination ($i$) of Sz\,28 with our setup. However, an inclusion would only enlarge the retrieved emitting areas by $\cos i$.
All dust species listed in Sect~\ref{sec:duckling} are included with sizes of $0.1\,\rm \mu m$, $2\,\rm \mu m$, and $5\,\rm \mu m$ in the retrieval.

The full Bayesian run with all molecules optimises $65$ parameters by UltraNest and further $31$ linear parameters by NNLS. This run finished after about $39\,\rm million$ model evaluations and approximately $1.64\,\rm days$ on $64$ CPUs ($\sim105$ CPU days).

After that, we exclude every molecule individually for a new Bayesian analysis. By calculating the logarithm of the Bayes factors ($\ln{B}$) from two runs, we can quantify the evidence for the presence of a molecule in the spectrum. The Bayes factor compares the global evidence for different fits. Values of $\ln{B}<1$, $1<\ln{B}<2.5$, $2.5<\ln{B}<5$, $5<\ln{B}<11$, and $11<\ln{B}$ can be translated to no evidence, weak evidence, moderate evidence, strong evidence, and very strong evidence, respectively, for one model over another model \citep{Trotta2008}. The sign of $\ln{B}$ indicates which model is preferred. A negative value indicates that the variant model is preferred whereas a positive value of $\ln{B}$ indicates the same for the original model. The Bayes factor is always calculated between the retrieval including all molecules and the one excluding a single molecule at a time.

\begin{table}[t]
    \centering
    \caption{Bayes factors that quantify the evidence for different molecules in the JWST/MIRI spectrum of Sz\,28.}
    \label{tab:bayes_mols}
    
\begin{tabular}{l|l|l|l}
\hline \hline&&&\\[-1.9ex]
mol\tablefootmark{(1)}  & $\ln{B}$\tablefootmark{(2)} & Pref\tablefootmark{(3)} & Evidence\tablefootmark{(4)} \\ \hline
&&&\\[-1.9ex]
 \ce{H2O} & -1.95 & no & weak \\ 
 \ce{NH3} & 8.18 & yes & strong \\ 
 \ce{CO} & 0.78 & yes & none \\ 
 \ce{C2H4} & -0.16 & no & none \\ 
 \ce{C2H2} & 1209.74 & yes & very strong \\ 
 \ce{CH4} & 161.37 & yes & very strong \\ 
 \ce{C2H6} & 38.12 & yes & very strong \\ 
 \ce{C3H4} & 11.47 & yes & very strong \\ 
 \ce{C4H2} & 72.49 & yes & very strong \\ 
 \ce{C6H6} & 184.58 & yes & very strong \\ 
 \ce{HC3N} & 16.50 & yes & very strong \\ 
 \ce{CH3} & 0.99 & yes & none \\ 
 \ce{CO2} & 63.94 & yes & very strong \\ 
 \ce{HCN} & 225.48 & yes & very strong \\ 
\end{tabular}
    \tablefoot{\\
\tablefoottext{1}{Excluded molecule compared to the original fit.}\\
\tablefoottext{2}{Logarithm of the Bayes factor between the fit with and without a molecule. Positive values indicate a preference for including this molecule.}\\
\tablefoottext{3}{Is there a preference for including this molecule?}\\
\tablefoottext{4}{Interpretation of $\ln B$ based on \cite{Trotta2008}.}
}
\end{table}

While there is very strong evidence for most species, as seen in Table~\ref{tab:bayes_mols}, we examine the few cases for which the evidence is less strong. 
These species are \ce{H2O}, \ce{NH3}, \ce{CO}, \ce{C2H4}, and \ce{CH3}. 

The negative value of $\ln{B}$ for \ce{H2O} indicates that the extra parameters for water are not worth the change in fit quality. The value of $-1.95$ indicates weak evidence against water. This is consistent with previous studies \citep{Arabhavi2024,Kanwar2024} that find a hydrocarbon-rich chemistry around VLMS that is depleted in oxygen. We discuss a possible upper limit for the \ce{H2O} abundance in Sz\,28 in Sect.~\ref{sec:water_limit}.

\begin{figure}
    \centering
    \includegraphics[width=\linewidth]{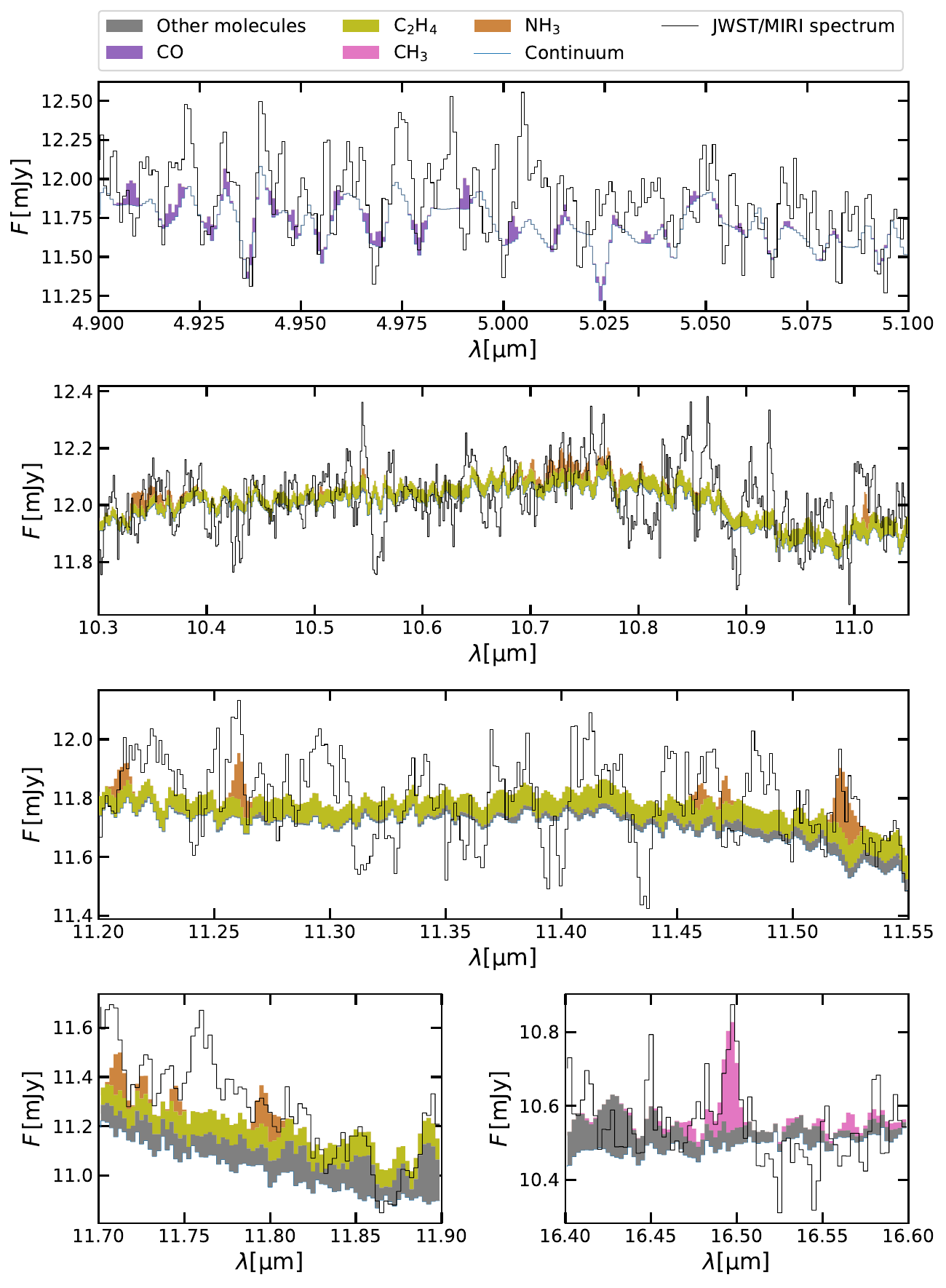}
    \caption{Zoom in on the molecular features of the maximum likelihood model for \ce{CO}, \ce{C2H4}, \ce{CH3}, and \ce{NH3}. The upper panel shows the wavelength region from $4.9\,\rm \mu m$ to $5.1\,\rm \mu m$, where \ce{CO} potentially emits. The middle ranks and lower left panel display three regions that exhibit \ce{NH3} emission and a \ce{C2H4} quasi-continuum. At the lower right panel (from $16.4\,\rm \mu m$ to $16.6\,\rm \mu m$) \ce{CH3} emission can be seen. The black line shows the JWST/MIRI spectrum of Sz\,28 with the blue line indicating the continuum, visible where no molecular emission is present.}
    \label{fig:weak_mols}
\end{figure}

There is no evidence for or against the presence of \ce{C2H4} in Sz\,28 ($\ln B = -0.16$). Analysing the retrieval including \ce{C2H4}, we cannot identify a \ce{C2H4} feature in the observation. The fit opted therefore for optically thick emission instead (middle and lower left panels of Fig.~\ref{fig:weak_mols}). This is reflected in the retrieved column density range from $8\times 10^{22} \,\rm cm^{-2}$ to $3\times 10^{23} \,\rm cm^{-2}$. Due to the combination of the inconclusive Bayes factor and the absence of a clear feature in model and observation, we see no evidence for \ce{C2H4} in Sz\,28.

\ce{CH3} produces non-significant evidence as well. As seen in the lower right panel of Fig.~\ref{fig:weak_mols}, \ce{CH3} is predicted to produce a clear feature (the Q-branch) at about $16.48\,\rm \mu m$. The observed weakness of that feature ($\sim 0.3\,\rm mJy$) explains why the Bayes factor is non-significant regarding this molecule. 
Additional lines of \ce{CH3} are too weak to be detected, with the lines of the P-branch that get stronger at high column densities falling beyond the analysed wavelength region (e.g. $17.6\,\rm \mu m$).
However, due to the strong visual confirmation of this molecule's Q-branch, we include it in the further study.

The Bayes factor of \ce{CO} ($\ln B = 0.78$) indicates that the inclusion of \ce{CO} does not improve the fit quality significantly. As shown by \cite{Kanwar2024}, the wavelength region that shows potential \ce{CO} emission is dominated by the stellar spectrum (upper panel of Fig.~\ref{fig:weak_mols}), which shows photospheric \ce{CO} absorption lines.  Therefore, our fit depends on the quality and sufficient resolution of the photospheric input spectrum. As seen in Fig.~\ref{fig:weak_mols}, there is no convincing evidence for \ce{CO} emission in Sz\,28 and a better description of the stellar spectrum is needed to draw further conclusions.

Lastly, the Bayes factor of \ce{NH3} ($\ln B = 8.18$) denotes strong evidence for the molecule emitting in Sz\,28. Analysing the spectral features (Fig.~\ref{fig:weak_mols}), there is some overlap of \ce{NH3} features with the JWST/MIRI spectrum (e.g. the features at about $11.52\,\rm \mu m$). However, the region is dominated by unexplained molecular features, which means that molecules that are not included in this retrieval or not included lines of examined molecules might explain the potential \ce{NH3} features as well. This can be especially seen in the second row of Fig.~\ref{fig:weak_mols}. Additionally, we retrieve very low temperatures for \ce{NH3} (from $99^{+50}_{-24}\, \rm K$ to $63^{+28}_{-13}\, \rm K$) together with rather high column densities (from $4^{\times6.2}_{\div14}(+19)\, \rm cm^{-2}$ to $8^{+60}_{-7}(+19)\, \rm cm^{-2}$). These low temperatures and high column densities are needed to reduce the strength of the features at $10.35\,\rm \mu m$ and $10.75\,\rm \mu m$ and increase the strength at $11.26\,\rm \mu m$ and $11.52\,\rm \mu m$. We argue that all the unaccounted molecular features lead to a higher dust continuum in our retrieval. A lower continuum might allow for stronger features at $10.35\,\rm \mu m$ and $10.75\,\rm \mu m$  and therefore different parameter values. All in all, the statistical evidence for \ce{NH3} is not corroborated by visual inspection of the spectrum. This means that some evidence for \ce{NH3} exists but further analysis is needed to claim the first detection in a JWST/MIRI spectrum. We include \ce{NH3} in the further analysis due to the improvement of the fitting it provides, keeping in mind that it does not overlap strongly with other included species (as stated above  \ce{C2H4} is excluded) and therefore has only a limited effect on their retrieved parameters.

Building on the results explained above, we exclude \ce{H2O}, \ce{CO}, and \ce{C2H4} from the further retrieval so that only the molecules with significant evidence are present. 
We note that excluding molecules from the fitting procedure will decrease the uncertainties of remaining molecules with overlapping wavelength regions with excluded molecules. However, comparing the posterior of the initial retrieval and the one excluding the species without significant evidence, we do not see a substantial change in retrieved parameter values and uncertainties. This hints that there are little degeneracies between the excluded and included parameters. Only the values for \ce{C2H6} change slightly, due to the molecule emitting in a similar wavelength range as the excluded \ce{C2H4}.

\subsection{Evidence for isotopologues\label{sec:iso}}

\cite{Kanwar2024} detected \ce{^{13}CCH2} and \ce{^{13}CO2} in the JWST/MIRI spectrum of Sz\,28.
We aim to evaluate the evidence for four isotopologues (\ce{^{13}CH4}, \ce{^{13}CCH6}, \ce{^{13}CCH2}, and \ce{^{13}CO2}), similarly to the evidence for different molecules as shown in Sect.~\ref{sec:mol_select}. 
The rarer isotopologues are not introduced as separate component, but rather as emitting under the same conditions as the main isotopologue. Therefore, they introduce no additional parameters. This is a simplification, since they are typically optically thinner, which means that deeper layer of the disk are contributing to their emission. However, this simplified treatment can account for line overlap between the main and rarer isotopologue since they are calculated in the same slab model. \ce{^{13}CH4} and \ce{^{13}CO2} are included with a ratio of $1:70$ to the main isotopologue, with \ce{^{13}CCH2} and \ce{^{13}CCH6} having a ratio of $1:35$ \citep{Woods2009}.

To quantify the evidence for isotopologues, we compare a Bayesian retrievals with and without the \ce{^{13}C} isotope. For the full run, \ce{^{13}CH4}, \ce{^{13}CCH6}, \ce{^{13}CCH2}, and \ce{^{13}CO2} were included under the same conditions as \ce{CH4}, \ce{CCH6}, \ce{CCH2}, and \ce{CO2}, respectively. All priors are identical to the ones presented in Sect.~\ref{sec:mol_select}, with the excluded molecules not being reintroduced.

\begin{table}[t]
    \centering
    \caption{Bayes factors that quantify the evidence for different isotopologues in the JWST/MIRI spectrum of Sz\,28.}
    \label{tab:bayes_isos}
\begin{tabular}{l|l|l|l}
\hline \hline&&&\\[-1.9ex]
mol\tablefootmark{(1)}  & $\ln{B}$\tablefootmark{(2)} & Pref\tablefootmark{(3)} & Evidence\tablefootmark{(4)} \\ \hline
&&&\\[-1.9ex]
 \ce{^13CH4} & 16.94 & yes & very strong \\ 
 \ce{^13CO2} & 5.20 & yes & strong \\ 
 \ce{^13CCH2} & 59.96 & yes & very strong \\ 
 \ce{^13CCH6} & -4.68 & no & moderate \\ 
\end{tabular}
    \tablefoot{\\
\tablefoottext{1}{Excluded molecule compared to the original fit.}\\
\tablefoottext{2}{Logarithm of the Bayes factor between the fit with and without an isotopologue. Positive values indicate a preference for including this isotopologue.}\\
\tablefoottext{3}{Is there a preference for including this isotopologue?}\\
\tablefoottext{4}{Interpretation of $\ln B$ based on \cite{Trotta2008}.}
}
\end{table}

The Bayes factors evaluating the evidence for the different isotopologues are listed in Table~\ref{tab:bayes_isos}. There is very strong evidence for \ce{^{13}CH4} and \ce{^{13}CCH2} and strong evidence for \ce{^{13}CO2}. 

\begin{figure}[t]
    \centering
    \includegraphics[width=\linewidth]{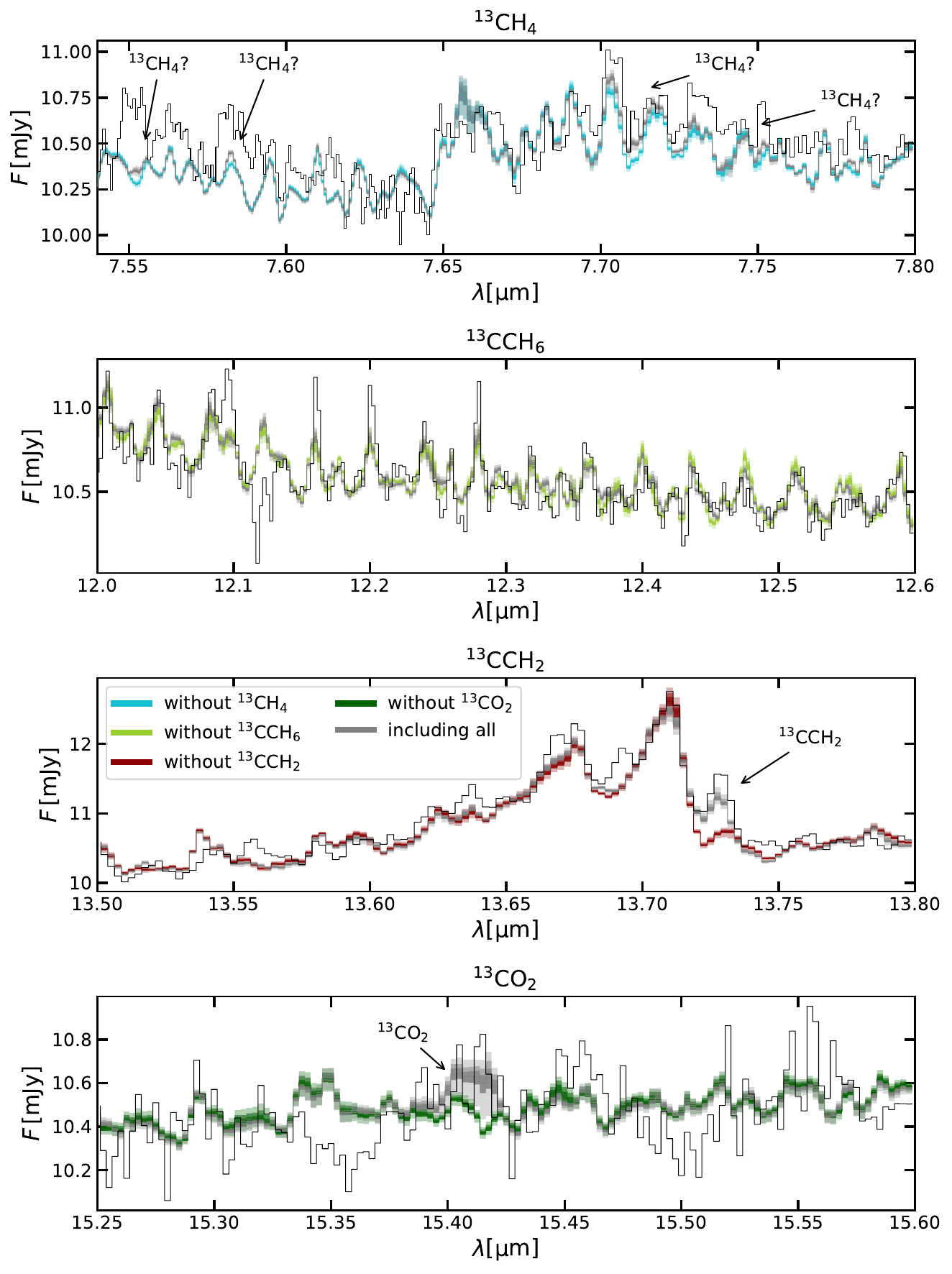}
    \caption{Zoom in on the isotopologue features comparing the version with all isotopologues (grey) with the ones excluding \ce{^{13}CH4}, \ce{^{13}CCH6}, \ce{^{13}CCH2}, and \ce{^{13}CO2}, respectively. While the JWST/MIRI spectrum is shown in $1\sigma$, $2\sigma$, and $3\sigma$ flux levels of the posteriors are displayed in lighter growing shades of their respective colours.}
    \label{fig:iso_evidence}
\end{figure}

The spectral differences can be seen in Fig.~\ref{fig:iso_evidence}. The upper panel shows the differences between the retrieval with (grey) and without \ce{^{13}CH4} (cyan). The Q-branch of \ce{^{13}CH4} is at about $7.7\,\rm \mu m$. Similarly, a few of wavelength regions (e.g. at about $7.55\,\rm \mu m$, $7.57\,\rm \mu m$ and $7.74\,\rm \mu m$) exhibit minor differences between both versions. The Bayes factor ($\ln B = 16.94$) reflects that all of the changes due to including \ce{^{13}CH4} improve the fitting quality (the posterior is moved closer to the observation).
However, the poor fitting quality in this wavelength range means that no clear \ce{^{13}CH4} are identifiable in the JWST/MIRI spectrum. Since the number of parameters stays the same regardless of isotopologues being included, we include \ce{^{13}CH4} in the further analysis since it improves the fit quality, but note that we do not claim a detection of this isotopologue.

For \ce{^{13}CCH2} and \ce{^{13}CO2}, there are clear features at $13.73\,\rm \mu m$ (third panel) and $15.41\,\rm \mu m$ (lowest panel) that are associated with the isotopologue, respectively. The versions excluding the respective isotopologues (burgundy and dark green) are therefore not able to reproduce the observed spectrum at these wavelengths.

The negative logarithm of the Bayes factor for \ce{^{13}CCH6} suggests that the exclusion of the isotopologue results in better overlap between the posterior of models and the observed spectrum. As seen in the second panel of Fig.~\ref{fig:iso_evidence}, the posterior without \ce{^{13}CCH6} (light green) shows more pronounced features. This is because most \ce{^{13}CCH6} features are in between the features of the main isotopologue. This `smoothed' spectrum is according to the Bayesian retrieval a better match (even though this is visually hard to confirm) to the observation (e.g. at $12.33\,\rm \mu m$ and $12.58\,\rm \mu m$). However, \ce{^{13}CCH6} might emit under different conditions than \ce{^{12}CCH6} which provides an alternative explanation of the Bayes factor next to the absence of \ce{^{13}CCH6}.

Based on these results we exclude the \ce{^{13}CCH6} from the analysis and derive the final set of molecules and isotopologues for which we find evidence in the retrieval.

\subsection{Emission conditions in Sz\,28 \label{sec:final_fit}}

The final retrieval incorporates the following molecules: \ce{C2H2}, \ce{^{13}CCH2}, \ce{HCN}, \ce{C6H6}, \ce{CO2}, \ce{^{13}CO2}, \ce{HC3N}, \ce{C2H6}, \ce{C3H4}, \ce{C4H2}, \ce{CH3}, \ce{CH4},\ce{^{13}CH4}, and \ce{NH3}. The isotopologue ratios are given in Sect.~\ref{sec:iso}. 
This results in $53$ parameters optimised by UltraNest and $28$ linear parameters.

\begin{figure*}[p]
    \centering
    \includegraphics[width=0.97\linewidth]{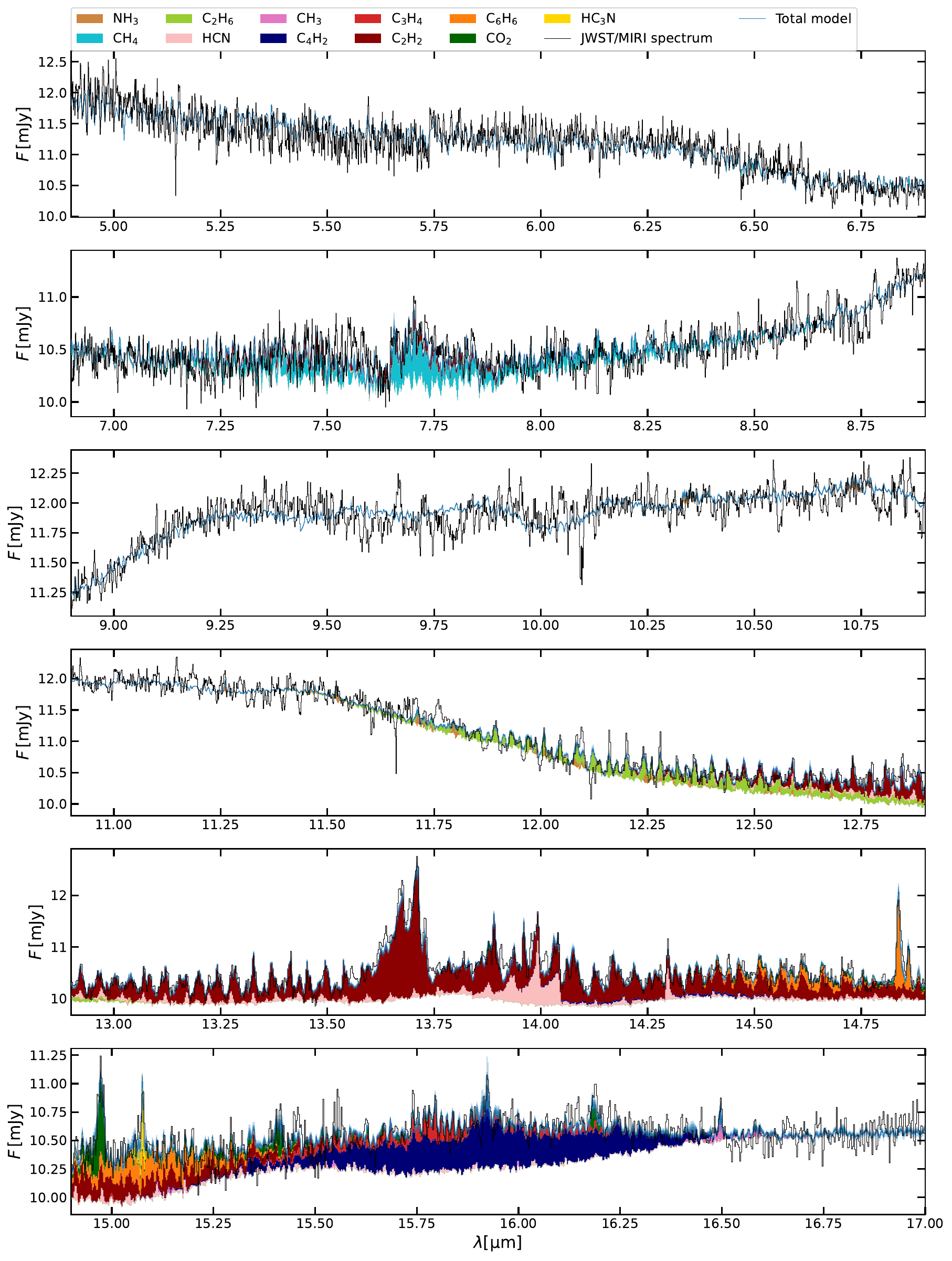}
    \caption{Molecular emission of the maximum likelihood model of Sz\,28. The JWST/MIRI spectrum is shown in black. The median flux of the posterior models is shown in blue with the $1\sigma$, $2\sigma$, and $3\sigma$ flux levels being displayed in lighter colours. The cumulative contributions of all included molecules are shown in different colours.}
    \label{fig:mol_spectrum}
\end{figure*}

Figure~\ref{fig:mol_spectrum} shows the posterior of model fluxes (blue line) and the maximum likelihood model from the posterior distribution (coloured areas). 
The molecular features of all molecules are clearly identifiable. The wavelength region (from about $8.5\,\rm \mu m$ to $11.5\,\rm \mu m$) between the \ce{CH4} feature and the line-rich region displays many unidentified molecular lines next to the potentially detected weak \ce{NH3} features. \ce{C2H4} has been identified at this wavelength region in ISO-ChaI-147 to form a quasi-continuum \citep{Arabhavi2024}. However, the \ce{CH2} wagging mode ($\nu7$) of \ce{C2H4} at $10.53\,\rm \mu m$ is not present in the JWST/MIRI spectrum of Sz\,28, and the species is not found in the spectrum according to the Bayes factor. Therefore, we speculate that other molecules not present in this analysis or missing lines of included molecules might be responsible for these features. \cite{Kanwar2024} provide a list of hydrocarbons that are chemically expected, but for which no mid-infrared spectral data is available in the HITRAN \citep{Gordon2022} and GEISA \citep{Delahaye2021} data collection. These could be candidates for the unaccounted features between $8.5\,\rm \mu m$ and $11.5\,\rm \mu m$.

The line-rich region longwards of $\sim11.5\,\rm \mu m$ is dominated by strong \ce{C2H2} features. We identify a quasi-continuum of \ce{C4H2} between about $15.5\,\rm \mu m$ and $16.5\,\rm \mu m$, which differs from the results by \cite{Kanwar2024}. The previous study fitted this region with a stronger dust continuum and postulated three scenarios to explain potential mismatches (we investigate this in Sect.~\ref{sec:quasi-continuum}). At wavelengths longward of the \ce{CH3} feature (at about $16.5\,\rm \mu m$), there is little molecular emission in the model that explains the observed spectral lines. We attribute this to missing molecules and missing lines in the spectral data of included molecules (e.g. missing lines of \ce{CH3}).

\begin{figure}
    \centering
    \includegraphics[width=\linewidth]{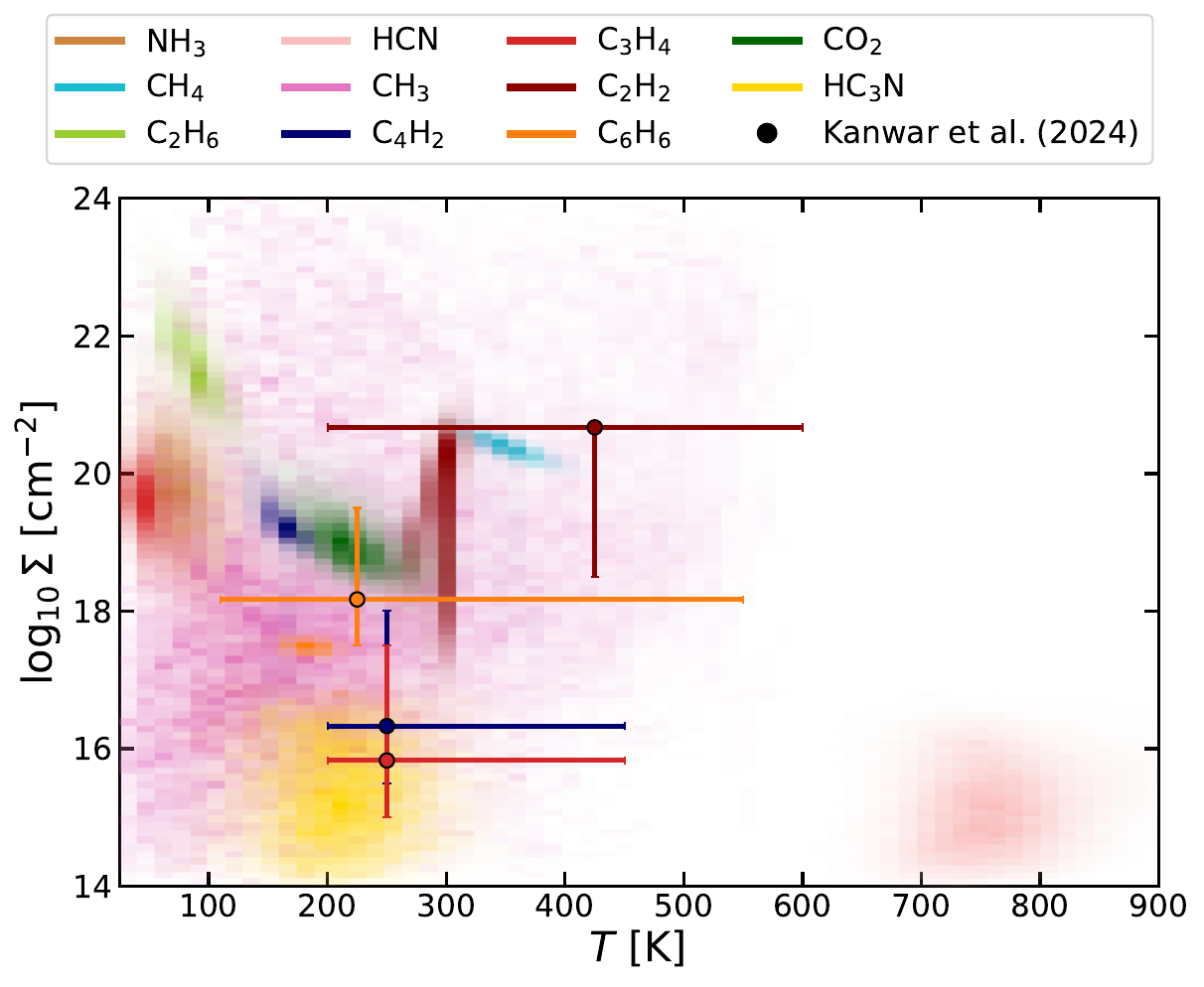}
    \caption{Molecular emission properties of Sz\,28. The different coloured areas denote where the respective molecule emits significantly in the posterior of models. The errorbars indicate the molecular emission conditions determined by \cite{Kanwar2024}.
    }
    \label{fig:mol_conditions}
\end{figure}

Next, we focus on the retrieved molecular and dust temperature properties which are listed with their uncertainties in Table~\ref{tab:sz28_posterior}. The relevant radial emitting area of all the molecules is determined by the central $70\,\%$ of radial flux emission, as described by \cite{Kaeufer2024}. The parameters with a $0.15$ and $0.85$ subscript denote the inner and outer edges of this region. The emitting areas are provided as the radii of disks with the same area. We calculate the radii ($r_{\rm eff}$) using the inner ($r_{0.15}$) and outer $r_{0.85}$ radius of significant emission:
\begin{align}
    r_{\rm eff}^2=r_{0.85}^2-r_{0.15}^2
\end{align}
The molecular properties are visualised in Fig.~\ref{fig:mol_conditions} with the slab parameters from \cite{Kanwar2024} shown for comparison.
\cite{Kanwar2024} fitted \ce{C2H2}, \ce{C6H6}, \ce{C4H2}, and \ce{C3H4} and detected \ce{CH4}, \ce{C2H6}, \ce{HCN}, \ce{CH3}, \ce{CO2}, and \ce{HC3N}.  
It is worth noting that the differences between both studies arise from the different fitting strategies \citep[e.g. the wavelength windows and sequential fitting employed by ][]{Kanwar2024} and from the new data reduction of the JWST/MIRI spectrum (Sect.~\ref{sec:obs}).

\begin{table}[t]
    \caption{Posterior parameter values and uncertainties for selected parameters of the Sz\,28 fit.}
    \centering
    \vspace{-2mm}
    \resizebox{!}{\linewidth}{
    \begin{tabular}{p{0.6cm} l|l|p{0.6cm}l|l}
\hline \hline & & & & & \\[-1.9ex] 
\multicolumn{2}{l|}{Parameter} & Posterior & \multicolumn{2}{l|}{Parameter} & Posterior \\ \hline  
  & & & & & \\[-1.9ex] 
\multicolumn{2}{c|}{$T_{\rm rim}$} & $974^{+11}_{-9}\, \rm K$ & &$r_{\rm eff}$ & $26^{+11}_{-25}\, \rm au$\\  
 & & & & & \\[-1.9ex] 
\multicolumn{2}{c|}{$T^{\rm sur}_{\rm min}$} & $292^{+13}_{-21}\, \rm K$ & &$t_{0.85}$ & $42^{+33}_{-12}\, \rm K$\\  
 & & & & & \\[-1.9ex] 
\multicolumn{2}{c|}{$T^{\rm sur}_{\rm max}$} & $800^{+500}_{-400}\, \rm K$ & $\rm C_3H_4$&$t_{0.15}$ & $64^{+60}_{-14}\, \rm K$\\  
 & & & & & \\[-1.9ex] 
\multicolumn{2}{c|}{$T^{\rm mid}_{\rm min}$} & $229^{+10}_{-11}\, \rm K$ & &$\Sigma_{0.85}$ & $5^{\times3.4}_{\div20}(+19)\, \rm cm^{-2}$\\  
 & & & & & \\[-1.9ex] 
\multicolumn{2}{c|}{$T^{\rm mid}_{\rm max}$} & $980^{+320}_{-330}\, \rm K$ & &$\Sigma_{0.15}$ & $7^{\times6.2}_{\div36}(+18)\, \rm cm^{-2}$\\  
 & & & & & \\[-1.9ex]\cline{4-5} 
 
\multicolumn{2}{c|}{$q_{\rm mid}$} & $-0.19^{+0.05}_{-0.05}$ & &$r_{\rm eff}$ & $0.058^{+0.005}_{-0.004}\, \rm au$\\  
 & & & & & \\[-1.9ex] 
\multicolumn{2}{c|}{$q_{\rm sur}$} & $-0.17^{+0.05}_{-0.11}$ & &$t_{0.85}$ & $286^{+10}_{-27}\, \rm K$\\  
 & & & & & \\[-1.9ex] 
\multicolumn{2}{c|}{$q_{\rm emis}$} & $-0.136^{+0.024}_{-0.11}$ & $\rm C_2H_2$&$t_{0.15}$ & $306^{+5}_{-6}\, \rm K$\\  
 & & & & & \\[-1.9ex] 
\multicolumn{2}{c|}{$a_{\rm obs}$} & $0.01436^{+0.00010}_{-0.00011}$ & &$\Sigma_{0.85}$ & $5^{+14}_{-4}(+17)\, \rm cm^{-2}$\\  
 & & & & & \\[-1.9ex] 
\cline{1-2} 
&$r_{\rm eff}$ & $14^{+80}_{-13}\, \rm au$ &&$\Sigma_{0.15}$ & $4.3^{+3.4}_{-1.3}(+20)\, \rm cm^{-2}$\\  
 & & & & & \\[-1.9ex]\cline{4-5} 
 
&$t_{0.85}$ & $64^{+40}_{-13}\, \rm K$ &&$r_{\rm eff}$ & $0.110^{+0.06}_{-0.032}\, \rm au$\\  
 & & & & & \\[-1.9ex] 
$\rm NH_3$&$t_{0.15}$ & $110^{+80}_{-40}\, \rm K$ &&$t_{0.85}$ & $172^{+16}_{-32}\, \rm K$\\  
 & & & & & \\[-1.9ex] 
&$\Sigma_{0.85}$ & $5^{+33}_{-4}(+19)\, \rm cm^{-2}$ &$\rm C_6H_6$&$t_{0.15}$ & $198^{+27}_{-15}\, \rm K$\\  
 & & & & & \\[-1.9ex] 
&$\Sigma_{0.15}$ & $2.6^{+18}_{-2.5}(+19)\, \rm cm^{-2}$ &&$\Sigma_{0.85}$ & $3.1^{+1.0}_{-0.8}(+17)\, \rm cm^{-2}$\\  
 & & & & & \\[-1.9ex] 
\cline{1-2} 
&$r_{\rm eff}$ & $0.026^{+0.008}_{-0.004}\, \rm au$ &&$\Sigma_{0.15}$ & $2.8^{+0.7}_{-0.6}(+17)\, \rm cm^{-2}$\\  
 & & & & & \\[-1.9ex]\cline{4-5} 
 
&$t_{0.85}$ & $330^{+21}_{-20}\, \rm K$ &&$r_{\rm eff}$ & $0.052^{+0.019}_{-0.013}\, \rm au$\\  
 & & & & & \\[-1.9ex] 
$\rm CH_4$&$t_{0.15}$ & $372^{+24}_{-22}\, \rm K$ &&$t_{0.85}$ & $203^{+29}_{-26}\, \rm K$\\  
 & & & & & \\[-1.9ex] 
&$\Sigma_{0.85}$ & $3.3^{+1.5}_{-1.0}(+20)\, \rm cm^{-2}$ &$\rm CO_2$&$t_{0.15}$ & $241^{+40}_{-31}\, \rm K$\\  
 & & & & & \\[-1.9ex] 
&$\Sigma_{0.15}$ & $1.4^{\times1.5}_{\div304}(+20)\, \rm cm^{-2}$ &&$\Sigma_{0.85}$ & $9^{+40}_{-6}(+18)\, \rm cm^{-2}$\\  
 & & & & & \\[-1.9ex] 
\cline{1-2} 
&$r_{\rm eff}$ & $1.7^{+6}_{-1.0}\, \rm au$ &&$\Sigma_{0.15}$ & $9^{+24}_{-6}(+18)\, \rm cm^{-2}$\\  
 & & & & & \\[-1.9ex]\cline{4-5} 
 
&$t_{0.85}$ & $78^{+17}_{-20}\, \rm K$ &&$r_{\rm eff}$ & $0.26^{+0.4}_{-0.16}\, \rm au$\\  
 & & & & & \\[-1.9ex] 
$\rm C_2H_6$&$t_{0.15}$ & $100^{+24}_{-16}\, \rm K$ &&$t_{0.85}$ & $190^{+50}_{-50}\, \rm K$\\  
 & & & & & \\[-1.9ex] 
&$\Sigma_{0.85}$ & $7^{+33}_{-5}(+21)\, \rm cm^{-2}$ &$\rm HC_3N$&$t_{0.15}$ & $230^{+50}_{-50}\, \rm K$\\  
 & & & & & \\[-1.9ex] 
&$\Sigma_{0.15}$ & $1.8^{+6}_{-1.4}(+21)\, \rm cm^{-2}$ &&$\Sigma_{0.85}$ & $1.6^{+11}_{-1.3}(+15)\, \rm cm^{-2}$\\  
 & & & & & \\[-1.9ex] 
\cline{1-2} 
&$r_{\rm eff}$ & $0.25^{+0.23}_{-0.14}\, \rm au$ &&$\Sigma_{0.15}$ & $3.6^{+23}_{-3.0}(+15)\, \rm cm^{-2}$\\  
 & & & & & \\[-1.9ex]\cline{4-5} 
 
&$t_{0.85}$ & $695^{+35}_{-40}\, \rm K$ &$\rm NH_3$&$\mathcal{N}_{\rm tot}$ & $7^{\times124}_{\div3039}(+48)$\\  
 & & & & & \\[-1.9ex]\cline{4-5} 
 
$\rm HCN$&$t_{0.15}$ & $820^{+70}_{-40}\, \rm K$ &$\rm CH_4$&$\mathcal{N}_{\rm tot}$ & $9^{+9}_{-5}(+43)$\\  
 & & & & & \\[-1.9ex]\cline{4-5} 
 
&$\Sigma_{0.85}$ & $8^{+40}_{-6}(+14)\, \rm cm^{-2}$ &$\rm C_2H_6$&$\mathcal{N}_{\rm tot}$ & $1.0^{\times72}_{\div16}(+49)$\\  
 & & & & & \\[-1.9ex]\cline{4-5} 
 
&$\Sigma_{0.15}$ & $1.9^{+8}_{-1.5}(+15)\, \rm cm^{-2}$ &$\rm HCN$&$\mathcal{N}_{\rm tot}$ & $5.1^{+0.4}_{-0.4}(+40)$\\  
 & & & & & \\[-1.9ex]\cline{4-5} 
 
\cline{1-2} 
&$r_{\rm eff}$ & $0.09^{+1.5}_{-0.08}\, \rm au$ &$\rm CH_3$&$\mathcal{N}_{\rm tot}$ & $1.1^{\times366}_{\div24}(+43)$\\  
 & & & & & \\[-1.9ex]\cline{4-5} 
 
&$t_{0.85}$ & $180^{+150}_{-100}\, \rm K$ &$\rm C_4H_2$&$\mathcal{N}_{\rm tot}$ & $3.5^{+17}_{-2.0}(+43)$\\  
 & & & & & \\[-1.9ex]\cline{4-5} 
 
$\rm CH_3$&$t_{0.15}$ & $220^{+170}_{-110}\, \rm K$ &$\rm C_3H_4$&$\mathcal{N}_{\rm tot}$ & $2.3^{\times4.1}_{\div37754}(+49)$\\  
 & & & & & \\[-1.9ex]\cline{4-5} 
 
&$\Sigma_{0.85}$ & $2.2^{\times1117}_{\div39}(+17)\, \rm cm^{-2}$ &$\rm C_2H_2$&$\mathcal{N}_{\rm tot}$ & $1.03^{+0.7}_{-0.24}(+44)$\\  
 & & & & & \\[-1.9ex]\cline{4-5} 
 
&$\Sigma_{0.15}$ & $3.1^{+1500}_{-3.0}(+18)\, \rm cm^{-2}$ &$\rm C_6H_6$&$\mathcal{N}_{\rm tot}$ & $2.4^{+4}_{-1.2}(+42)$\\  
 & & & & & \\[-1.9ex]\cline{4-5} 
 
\cline{1-2} 
&$r_{\rm eff}$ & $0.066^{+0.06}_{-0.020}\, \rm au$ &$\rm CO_2$&$\mathcal{N}_{\rm tot}$ & $1.8^{+6}_{-1.2}(+43)$\\  
 & & & & & \\[-1.9ex]\cline{4-5} 
 
&$t_{0.85}$ & $152^{+15}_{-17}\, \rm K$ &$\rm HC_3N$&$\mathcal{N}_{\rm tot}$ & $8^{+25}_{-4}(+40)$\\  
 & & & & & \\[-1.9ex] 
$\rm C_4H_2$&$t_{0.15}$ & $173^{+17}_{-15}\, \rm K$ & & \\ & & & & & \\[-1.9ex] 
&$\Sigma_{0.85}$ & $2.0^{+7}_{-1.5}(+19)\, \rm cm^{-2}$ & & \\ & & & & & \\[-1.9ex] 
&$\Sigma_{0.15}$ & $1.3^{+1.0}_{-0.6}(+19)\, \rm cm^{-2}$ & & \\
\end{tabular}}
\tablefoot{\\
The notation $a(+b)$ means $a\times10^b$. For parameters that have a lower limit of more than an order of magnitude below the retrieved value, the uncertainties are provided as factors. $\mathcal{N}_{\rm tot}$ denotes the total number of molecules that contributed to the described emission and not the total number of molecules in the disk.}
    \label{tab:sz28_posterior}
\end{table}

Focusing on the four molecules fitted by \cite{Kanwar2024}, we identify that the emission conditions of \ce{C2H2} and \ce{C6H6} overlap well within the error bars between both studies. However, we find lower temperatures and higher column densities for \ce{C4H2} and \ce{C3H4} compared to \cite{Kanwar2024}. This can be explained by the quasi-continuum of \ce{C4H2} that we identify in contrast to the previous study. We evaluate the robustness of this finding in Sect.~\ref{sec:quasi-continuum}.

Generally, the retrieved molecular conditions for some molecules are well constrained (e.g. \ce{CO2}, \ce{C6H6}, \ce{C4H2}, \ce{C2H6}, and \ce{CH4}). This means that there is little ambiguity in which molecular conditions are able to reproduce the observed spectrum. It also means that the emission is originating from relatively homogenous regions. On the other hand, some molecules (most prominently \ce{CH3}) have poorly constrained emission conditions. This is in line with our earlier conclusion that there is little evidence for \ce{CH3} in the observed spectrum which makes it more difficult to pin down the \ce{CH3} parameters. 

It strikes that \ce{HCN} is emitting at higher temperatures (from $695^{+35}_{-40}\,\rm K$ to $820^{+70}_{-40}\,\rm K$) than the rest of the molecules ($\lesssim 400\,\rm K$). This hints that \ce{HCN} is emitting in a different region of Sz\,28 which we further investigate in Sect.~\ref{sec:location}.

\ce{NH3}, \ce{C2H6}, and \ce{C3H4} show very low retrieved temperatures (upper limits of $110^{+80}_{-40}\,\rm K$, $100^{+24}_{-16}\,\rm K$, and $64^{+60}_{-14}\,\rm K$, respectively). To reproduce the observed features with these low-temperature models, high column densities and emitting areas are needed. This is reflected in the total number of emitting molecules listed in Table~\ref{tab:sz28_posterior}. While the posterior median molecule numbers for \ce{NH3}, \ce{C2H6}, and \ce{C3H4} are all about $10^{49}$, the $-1\sigma$ limits are $2\times 10^{45}$, $6.5\times 10^{47}$, $6\times 10^{44}$, respectively. This suggests that models with an extremely high number of molecules might explain the observation best, but warmer models with much fewer molecules are also possible. We tested this hypothesis by running another retrieval, limiting the emitting areas to a maximum radius of $2\,\rm au$. Consequently, the emitting areas decrease and temperatures rise (see Table~\ref{tab:posterior_restrict}). The restriction has no strong effect on the retrieved fit quality ($\ln B =4.98$, see also Fig.~\ref{fig:mol_spectrum_restricted}), showing that the huge emitting areas and low temperatures from a free retrieval are not the only options to explain the JWST/MIRI spectrum of Sz\,28.

The dust temperature structure is well constrained, except for the maximum temperature of the midplane and surface layer. We suspect that the maximum temperatures of both components have an insignificant effect on the flux in the observed wavelength region. This is why they are mostly constrained by their priors which can lead to the unphysical situations that the rim temperature is lower than the maximum temperature of midplane and surface layer. Therefore, these values are not analysed in detail. The minimum temperature of both components and the rim temperature are however well constrained (within about $10\,\rm K$). The power law indices for the midplane, surface, and molecular layer are on the shallow end of the prior range and within each other's uncertainties. Therefore, it is possible that the dust and molecular emission trace a similar disk layer, but a more in-depth analysis is needed to confirm this. Additionally, we note that the high dust temperatures compared to the temperatures of molecules like \ce{NH3}, \ce{C3H4}, and \ce{C2H6} mean that if interactions between the different model components would be included in the model, gas absorption features might appear (as discussed in Sect. \ref{sec:duckling}).

The derived observational uncertainty ($a_{\rm obs}$) of $1.436^{+0.010}_{-0.011}\,\%$ together with the mean observational flux of $11\,\rm  mJy$ translates into an uncertainty of $0.16\,\rm mJy$ which is slightly larger than the uncertainty of about $0.1\,\rm mJy$ estimated with the Exposure Time Calculator \citep{Kanwar2024}. This is expected since $a_{\rm obs}$ resembles the observed uncertainty that best explains the difference between models and observations. Therefore, the many unidentified lines between $9\,\rm \mu m$ and $12\,\rm \mu m$ increase $a_{\rm obs}$. Fixing the observed uncertainty to $1\,\%$ of the observed flux results in a nearly identical posterior distribution compared to the treatments of the uncertainty as a free parameter. Therefore, the retrieval of the uncertainty eliminates the need to fix this parameter without any significant change regarding the retrieved distribution for all other parameters.

\section{Discussion\label{sec:discussion}}

\subsection{Emission location\label{sec:location}}

\begin{figure}[t]
    \centering
    \includegraphics[width=\linewidth]{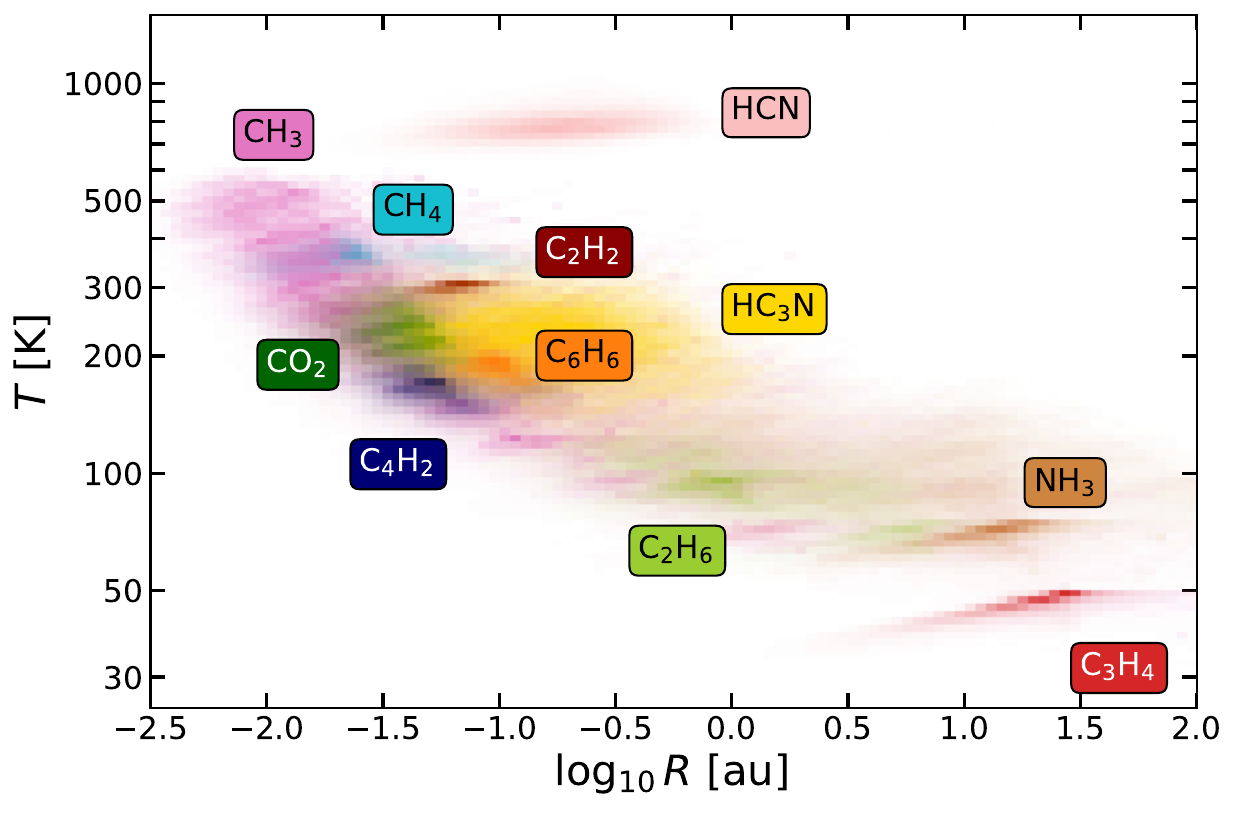}
    \caption{Radial molecular emission regions of Sz\,28. The different coloured areas denote where the respective molecule emits significantly in the posterior of models.}
    \label{fig:mol_conditions_radially}
\end{figure}

The molecular flux in DuCKLinG is described by a 1D radial structure of temperature and column density. After having discussed the temperature and column density ranges in the previous section, we now have a closer look at the radial regions from which the molecules emit.
As shown in Sect.~\ref{sec:duckling}, the emission of every molecule follows a radial power law in temperature and column density. The temperature decreases radially (due to the prior of $q_{\rm emis}$ in Table~\ref{tab:prior_full_fit}), but the column density can increase or decrease based on $\Sigma_{\rm tmax}$ and $\Sigma_{\rm tmin}$. Fig.~\ref{fig:mol_conditions_radially} indicates the regions from which the molecules emit significantly (between $t_{0.85}$ and $t_{0.15}$). 

We estimate the stellar radius of Sz\,28 based on pre-mainsequence evolutionary tracks from \cite{Siess2000}, an effective temperature of $3060\,\rm K$, and a luminosity of $0.04\, \rm L_{\odot}$ to be $0.0025\,\rm au$ which corresponds to a $log_{10}$ value of the radius of $-2.6$. Therefore, the stellar radius is at about the left edge of Fig.~\ref{fig:mol_conditions_radially}.

It can be seen that the emissions of all molecules except for \ce{HCN} form a radial power law in temperature. We stress that this is not an input to the model. We condition all molecules to have the same power law index ($q_{\rm emis}$), but the temperatures at the same radius are allowed to differ. The retrieved radial power law might indicate that the molecules are emitted from a similar layer of the disk instead of different depths at the same radius.

The emission of \ce{HCN} does not align with all other molecules. The high temperatures (from about $800\,\rm K$ to $700\,\rm K$) are present at radii from about $0.1\,\rm au$ to $1\,\rm au$. However, the column densities of about $10^{14}\,\rm cm^{-2}$ result in the lowest number of emitting molecules ($5.1^{+0.4}_{-0.4}(+40)$) of all examined species. The high temperatures and low column densities at relatively large radii suggest that \ce{HCN} is emitted in an upper layer of the disk. \cite{Woitke2024} find most of the emission of \ce{HCN} to originate close to the midplane in a thermo-chemical model describing the JWST/MIRI spectrum of Ex\,Lup. However, lower concentrations of the molecule are also present in a higher radially extended layer, which might explain the emission of \ce{HCN} in the case of Sz\,28. Earlier \textit{Spitzer} results also generally find high temperatures ($\gtrsim 700\,\rm K$) of \ce{HCN} in T\,Tauri and Herbig disks \citep{Salyk2011}.

As noted before, \ce{C2H6}, \ce{NH3}, and \ce{C3H4} emit at very low temperatures from large emission areas. As detailed in Sect.~\ref{sec:final_fit}, the posteriors of these molecules extend to smaller emitting areas and limiting the emitting radii to less than $2\,\rm au$ does not significantly impact the fit quality.

\ce{CH3} is rather poorly constrained. Therefore, we are not analysing the radial behaviour in detail. 

\ce{CO2}, \ce{CH4}, \ce{C2H2}, \ce{C4H2}, and \ce{C6H6} emit at well-defined very similar radial regions. 
Therefore, we have good reasons to assume that these molecules are co-spatial in this disk. 
The mean temperature in a radial range between $0.02\,\rm au$ and $0.2\,\rm au$ of \ce{CO2} ($220\,\rm K$), \ce{C4H2} ($160\,\rm K$), and \ce{C6H6} ($180\,\rm K$) are slightly lower than the temperatures of \ce{CH4} ($350\,\rm K$) and \ce{C2H2} ($290\,\rm K$), suggesting that \ce{CO2} and \ce{C4H2} emit slightly deeper in the disk or in the case of \ce{C6H6} at slightly larger radii (Fig.~\ref{fig:mol_conditions_radially}).
\ce{HC3N} originates potentially from the same region but extends to larger radii. 
The full thermo-chemical disk models with ProDiMo for Sz\,28 by \cite{Kanwar2024} suggest that the inner disk must have a high $\rm C/O$ ratio to explain the large abundances of the hydrocarbon molecules. A chemically consistent solution to the presence of \ce{CO2} along with hydrocarbons can be a non-co-spatial origin in case of $\rm C/O\!\gg\!2$ \citep{Kanwar2024b}. 
However, for moderate $\rm C/O\!\approx\!2$ it can be co-spatial \citep{Kanwar2024}, and for close to solar $\rm C/O\!=\!0.46$ it can as well, see ProDiMo models for EX\,Lupi \citep{Woitke2024}.

In the latter model, the irradiation of the inner disk by X-rays unlocks the CO on timescales of a few years, and the liberated oxygen and carbon atoms react quickly to form \ce{C2H2} and HCN that are co-spatial with \ce{CO2}, although the line emission region for \ce{CO2} results to be situated at slightly larger radii compared to \ce{C2H2} and HCN, which is different from the results of our retrieval models presented in this paper. The ProDiMo models for both objects show that the emission lines from \ce{HCN} and \ce{C2H2} originate in similar disk layers, which is also different from the results found in this paper. 

\cite{Temmink2024} find that \ce{HCN}, \ce{C2H2}, and \ce{CO2} are emitting at increasing depths in the disk of the T\,Tauri star DR\,Tau. We can confirm this behaviour for the disk of Sz\,28, with the three molecules showing decreasing emitting temperatures at similar radii. 

\subsection{Constraining the water abundance in Sz\,28 \label{sec:water_limit}}

\begin{figure}[t]
    \centering
    \includegraphics[width=0.9\linewidth]{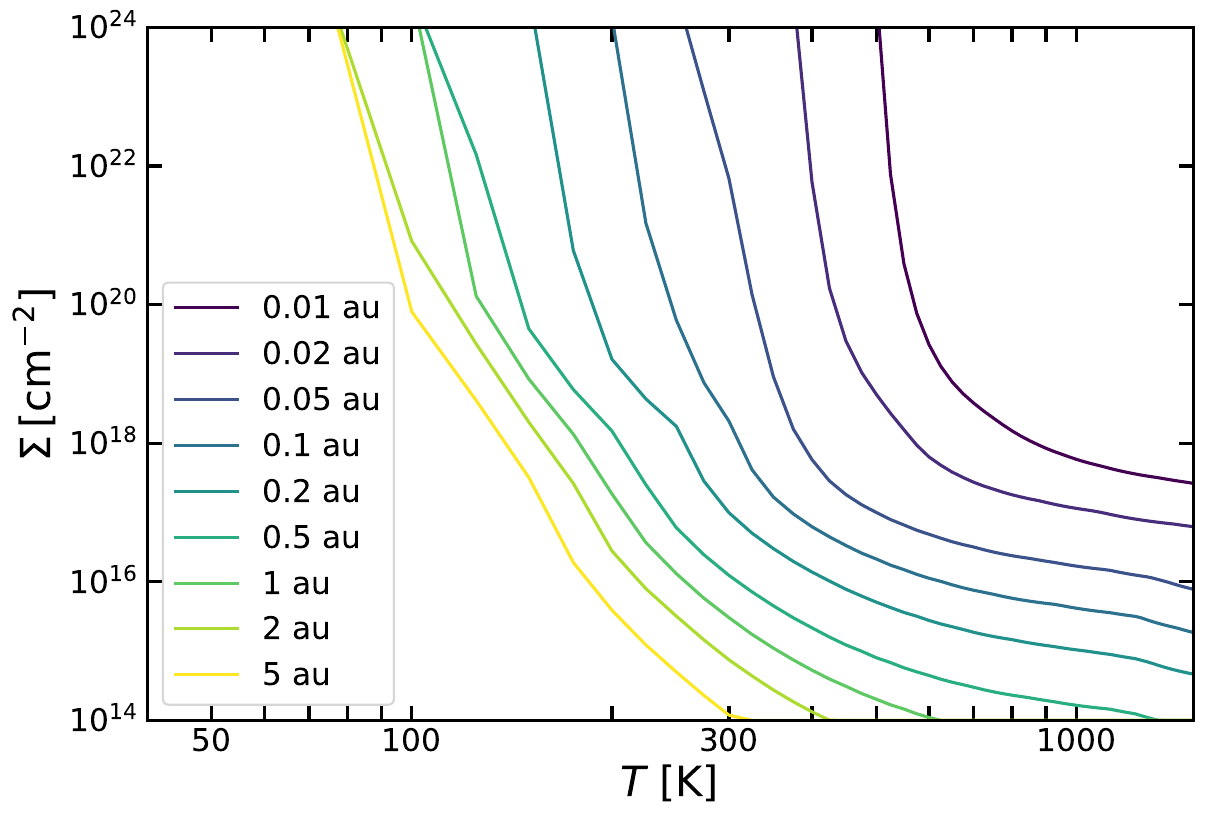}
    \caption{Column densities of water for given temperatures (horizontal axis) and emitting radii (colour) that result in peak fluxes of $3\sigma$ of the noise of the JWST/MIRI observation of Sz\,28.}
    \label{fig:water_limit}
\end{figure}

There is no evidence for water in the JWST/MIRI spectrum of Sz\,28 (according to the Bayes factor presented in Sect.~\ref{sec:mol_select}). This led us to determine an upper limit for the water abundance. We estimated this limit by selecting slab models from the grid presented by \cite{Arabhavi2024} that result in peak fluxes below $3\sigma$ of the noise level of Sz\,28 \citep[$0.1\,\rm mJy$,][]{Kanwar2024}.

Figure~\ref{fig:water_limit} shows the column densities of water for given temperatures and emitting areas (characterized by the radius of this area) that result in peak fluxes below $0.3\,\rm mJy$ in the fitted wavelength range of Sz\,28. 

At high temperatures ($\gtrsim 600\,\rm K$), when assuming large emitting areas ($r_{\rm eff}\gtrsim 1\,\rm au$), even the lowest examined column densities ($10^{14}\,\rm cm^{-2}$) are detectable for Sz\,28.
If the water emission temperature is only $200\,\rm K$, again assuming an emitting radius of $1\,\rm au$, the upper limit for the water column density is about $2\times10^{17}\,\rm cm^{-2}$.
This shows that small but substantial amounts of water could be hidden in the disk of Sz\,28 without producing noticeable signals. Previous findings for the VLMS J160532, that column densities of $3\times 10^{18}\,\rm cm^{-2}$ are not detectable if \ce{H2O} is emitting under the same conditions as \ce{C2H2}, support our conclusion \citep{Tabone2023}. This phenomenon of water as a weak emitter compared to hydrocarbons is examined in further detail by Arabhavi et al. (in prep). 

\subsection{Evidence for a \ce{C4H2} quasi-continuum\label{sec:quasi-continuum}}

The most significant difference in the retrieved molecular emission between our study and \cite{Kanwar2024} is the \ce{C4H2} quasi-continuum from about $15.5\,\rm \mu m$ to $16.4\,\rm \mu m$. \cite{Kanwar2024} proposed three scenarios to explain the mismatches at these wavelengths of their fit: missing dust species, incomplete opacity data, or a molecular quasi-continuum. Using the Bayesian approach in our study we can quantify the evidence for this quasi-continuum. We investigate the \ce{C4H2} properties in two ways. First, we show how limiting the integrated flux of \ce{C4H2} affects the retrieval. Secondly, we examine if the dust continuum at about $15.5\,\rm \mu m$ to $16.4\,\rm \mu m$ is a consequence of the continuum at shorter wavelengths or the best solution for the line-rich region from $\sim 12.0\,\rm \mu m$ onwards as well. This latter case would mean that fitting the wavelength region from $12\,\rm \mu m$ to $17\,\rm \mu m$ should result in the same quasi-continuum.

\begin{figure}
    \centering
    \includegraphics[width=0.9\linewidth]{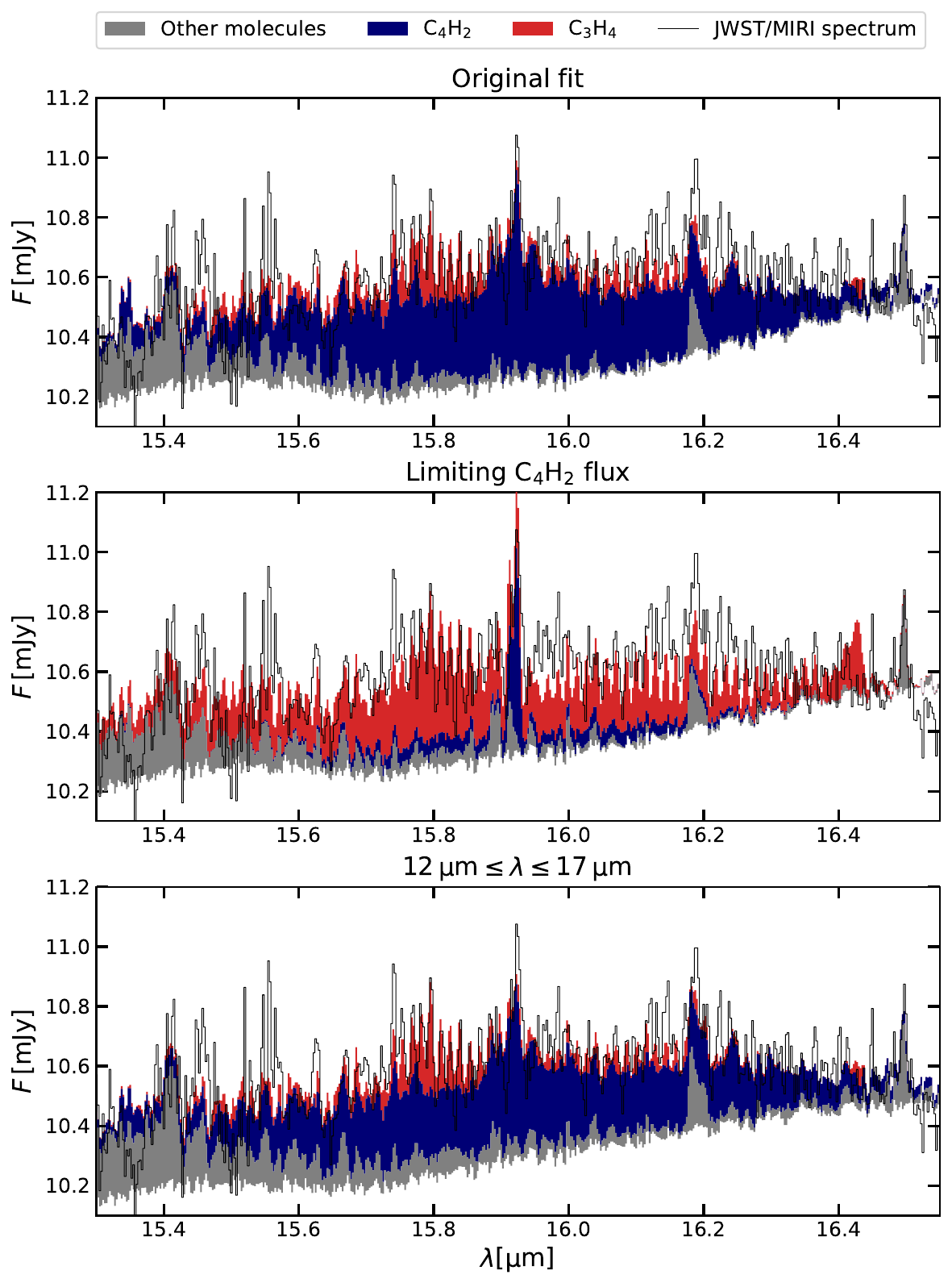}
    \caption{Zoom in to the \ce{C4H2} quasi-continuum of the maximum likelihood model for the retrieval presented in Sect.~\ref{sec:final_fit} (upper panel), the retrieval restricting the \ce{C4H2} flux (middle panel) and for the retrieval focusing only on the wavelength region from $12\,\rm \mu m$ to $17\,\rm \mu m$ (lower panel). The JWST/MIRI spectrum of Sz\,28 is shown in black, with the molecular contributions to the maximum likelihood model being shown in different colours.}
    \label{fig:c4h2_quasi}
\end{figure}

The first retrieval uses the same priors and settings as the final run presented in Sect.~\ref{sec:final_fit}. Additionally, we constrain the integrated flux of \ce{C4H2} to be below $4.5\times10^{-19} \,\rm W /m^2$ by setting the likelihood of higher integrated fluxes to highly unlikely values. This limit is about the \ce{C4H2} flux between $15.85\,\rm \mu m$ and $15.95\,\rm \mu m$ from the maximum likelihood model without any constraints. The restriction excludes the possibility of a \ce{C4H2} quasi-continuum while still allowing for \ce{C4H2} emission to describe the features at about $15.9\,\rm \mu m$. Analysing the newly retrieved parameters, we find that the values for \ce{C4H2} (column density range from $3.1^{+5}_{-1.9}(+16)\, \rm cm^{-2}$ to $4.5^{+17}_{-3.5}(+17)\, \rm cm^{-2}$ and a temperature range from  $210^{+80}_{-80}\, \rm K$ to $117^{+50}_{-33}\, \rm K$) fall within the retrieved values from \cite{Kanwar2024}. However, the temperature for \ce{C3H4} stays very low ($\lesssim 100\,\rm K$) with the column density reaching a maximum value of $8.6^{+1.1}_{-1.9}(+23)\, \rm cm^{-2}$. The middle panel of Fig.~\ref{fig:c4h2_quasi} shows the wavelength region of the \ce{C4H2} quasi-continuum for the new retrieval. It can be seen that limiting the molecular flux from \ce{C4H2} results in more flux from \ce{C3H4} with the dust continuum only increasing slightly in flux (comparing the upper and middle panel). However, this new fitting quality is significantly worse compared to the unconstrained fit. For example, there are large underpredictions in the wavelength region from $16.1\,\rm \rm \mu m$ to $16.3\,\rm \rm \mu m$. Additionally, \ce{C3H4} is producing a strong feature at about $16.4 \,\rm \mu m$ that is not mirrored in the observation. All of this is reflected in the logarithm of the Bayes factor of $58.82$ between the original/unconstrained and limited fit. Even though the Bayes factor for this exercise is positive by definition (same prior ranges, but a likelihood penalty for the limited fit), the large value indicates very strong evidence for the unconstrained fit. This makes us confident, that the quasi-continuum of \ce{C4H2} cannot easily be replaced by other molecular emissions or a higher dust continuum if examining the full wavelength range. 

To examine whether a more flexible dust continuum can replace the \ce{C4H2} quasi-continuum , we exclude the observations shortward of $12\,\rm \mu m$ from the analysis. There is theoretical \citep{Jang2024} and observational \citep{Kessler2006} evidence that the dust composition changes within a disk resulting in different wavelength regions probing different dust compositions. Since we do not account for radial changes in dust properties in our 1D model, focusing on small wavelength ranges is an alternative to allow for dust composition gradients. The lower panel of Fig.~\ref{fig:c4h2_quasi} illustrates the maximum likelihood model of the retrieval from $12\,\rm \mu m$ to $17\,\rm \mu m$ with the same priors and settings as mentioned in Sect.~\ref{sec:final_fit}. The resulting molecular fluxes are very similar to the one shown in Fig.~\ref{fig:mol_spectrum}. The retrieved parameters for \ce{C4H2} (column density from $1.6^{\times6.0}_{\div505}(+18)\, \rm cm^{-2}$ to $2.2^{+10}_{-1.9}(+19)\, \rm cm^{-2}$ with a temperature from $236^{+40}_{-33}\, \rm K$ to $183^{+32}_{-33}\, \rm K$) show a slight increase in temperature with a coinciding decrease in column density. Therefore, they are again closer to the values retrieved by \cite{Kanwar2024} but still show a quasi-continuum. This illustrates that the changes in retrieved conditions between both studies might partly be due to differences in the dust continuum. Summarizing, the persistence of a \ce{C4H2} quasi-continuum shows that it is not a consequence of a restricted dust continuum fit over the short wavelength region but indeed the best-fitting solution for this molecular line-rich wavelength region. 

Both of these tests lead to the same finding of a \ce{C4H2} quasi-continuum that is robust and not easily replaceable by other emissions or a change of continuum in our model. This nicely illustrates the benefit of a simultaneous fit of all molecules and the dust continuum.

\subsection{Dust composition\label{sec:dust_composition}}

\begin{figure}
    \centering
    \includegraphics[width=0.9\linewidth]{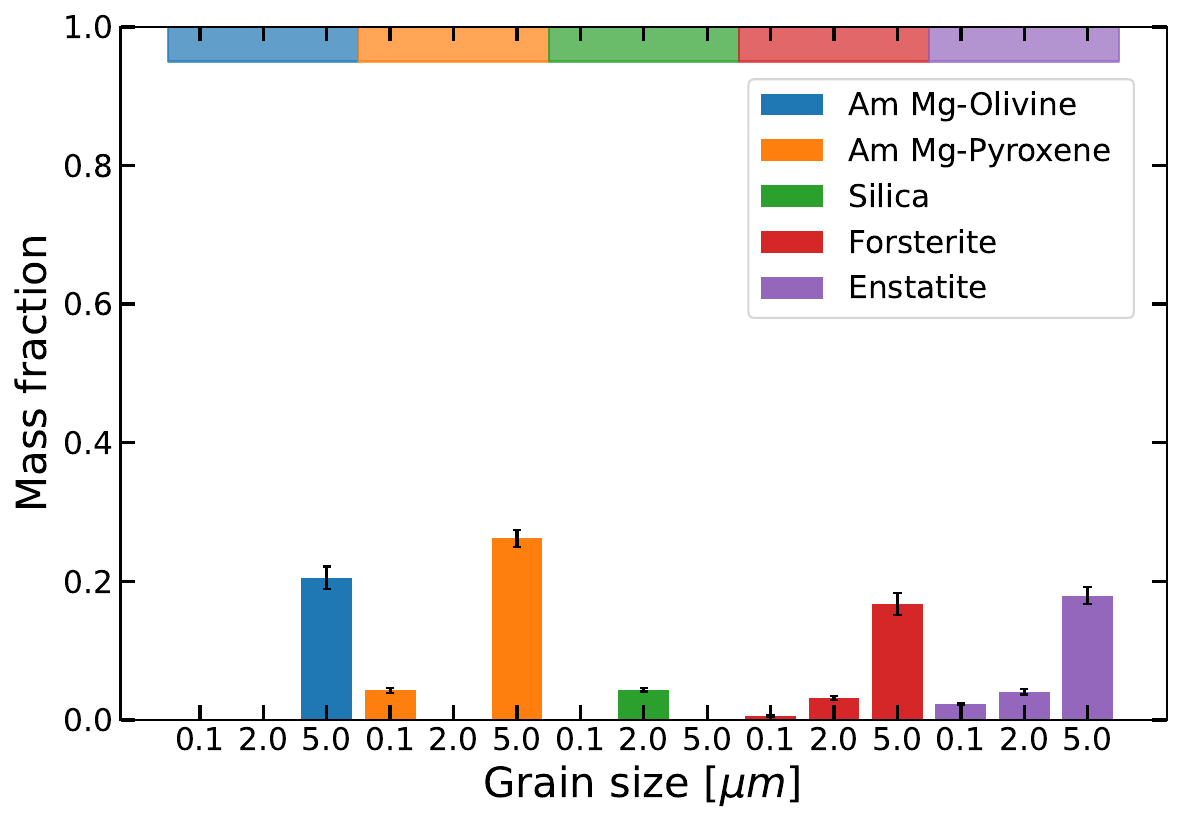}
    \caption{Mass fraction of the optically thin dust. The histograms indicate the median retrieved mass fraction of every dust species (composition and size), with the error bars denoting the $16$th and $84$th percentile. The dust composition is indicated by colour with the size displayed on the x-axis.}
    \label{fig:mass_fractions}
\end{figure}

Next to the gas properties, DuCKLinG allows for the retrieval of the dust composition. Fig.~\ref{fig:mass_fractions} illustrates the dust mass fraction for all species (composition and sizes) used in the fitting procedure. 

For all grain compositions (except for Silica), the largest grain size ($5\,\rm \mu m$) is the most common component. Additionally, $44.7^{+2.1}_{-2.1} \,\%$ of the optically thin dust is crystalline. Both factors indicate that the dust in Sz\,28 is highly evolved. The same conclusion is drawn for Sz\,28 by \cite{Kanwar2024}, even though the detailed numbers (crystalline fraction of $\sim 20\,\%$) differ. These minor differences can be due to e.g. the new data reduction of the JWST/MIRI spectrum but also the different wavelength regions considered in both studies. While \cite{Kanwar2024} fitted the dust continuum in two wavelength regions (from $4.9\,\rm \mu m$ to $15\,\rm \mu m$ and from $15\,\rm \mu m$ to $22\,\rm \mu m$) by an iterative process between dust and gas fitting, our study investigates the region from $4.9\,\rm \mu m$ to $17\,\rm \mu m$ with a single set of dust (and gas) parameters. As mentioned before, different wavelength regions probe the dust composition at different disk positions. We conclude from both studies that describing the dust with a single composition up to $17\,\rm \mu m$ is feasible, while larger wavelength regions require gradients in the dust compositions.

The crystallinity fraction of more than $40\,\%$ is consistent with findings for other VLMS \citep{Apai2005}. This high fraction is mostly driven by the abundances of forsterite and enstatite with grain sizes of $5\,\rm \mu m$. To test if these grains are needed to describe the observation, we run an additional retrieval excluding them. The Bayes factor ($ln B=36$) between the fit including all dust species and the one without forsterite and enstatite with grain sizes of $5\,\rm \mu m$ provides very strong evidence that these grains are indeed necessary to sufficiently describe the observation.

\begin{figure}
    \centering
    \includegraphics[width=0.9\linewidth]{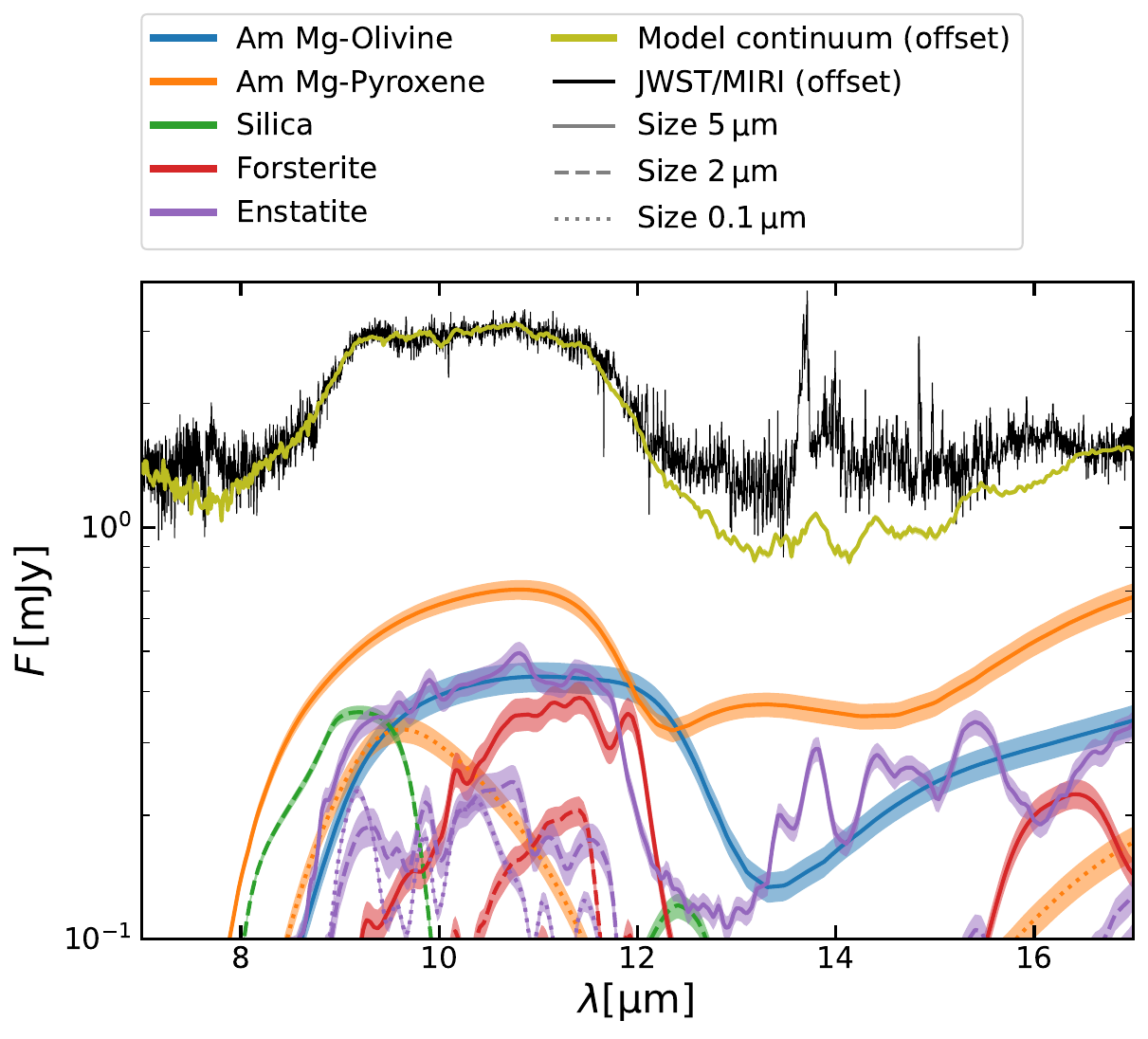}
    \caption{Spectrum of the different dust compositions (colour) and sizes (line style) compared to the total dust continuum (olive; star, rim, midplane, and surface layer) and the JWST/MIRI spectrum of Sz\,28 (black). Both are offset by $-9\,\rm mJy$. The error tubes denote the $16$th and $84$th percentile of the dust flux from the posterior of models.}
    \label{fig:dust_spectrum}
\end{figure}

Interestingly, we find very similar sizes and a similar ratio of pyroxene to olivine in the crystalline and amorphous silicate components.

The effect of the dust composition on the spectrum can be seen in Fig.~\ref{fig:dust_spectrum}. The grains of size $5\,\rm \mu m$ exhibit the strongest influence on the dust spectrum. The total continuum follows neatly the JWST/MIRI spectrum up to about $11.5\,\rm \mu m$. Longwards of that, the molecular emission contributes strongly to the total flux. This highlights again the difficulty disentangling the dust and gas contributions of JWST/MIRI spectra.

\section{Summary and conclusion\label{sec:summary}}

In this paper, we analysed the dust and gas composition of Sz\,28 simultaneously in a Bayesian way. This has been done using the 1D DuCKLinG model that describes the dust and gas at the same time. Sz\,28 is a VLMS exhibiting many molecular features on top of dust features which complicates the fitting process.
The main findings are listed below:

\begin{itemize}
    \item We confirm the previous identification by \cite{Kanwar2024} of \ce{C2H2}, \ce{HCN}, \ce{C6H6}, \ce{CO2}, \ce{HC3N}, \ce{C2H6}, \ce{C3H4}, \ce{C4H2}, \ce{CH4}, and \ce{CH3} in the disk of Sz\,28. We quantify the confidence in these detections as very strong evidence for all molecules except for \ce{CH3}. 
    \item There is no evidence for water emission in the JWST/MIRI spectrum of Sz\,28 up to $17\,\rm \mu m$. However, we estimate that it is possible to hide large quantities of \ce{H2O} in the noise of the spectrum. For example, at a temperature of $200\,\rm K$ assuming an emitting areas with a radius of $1\,\rm au$, the upper limit for the water column density is about $2\times10^{17}\,\rm cm^{-2}$.
    \item We find tentative hints for the presence of \ce{NH3}. While there is statistical evidence for this tentative detection, the wavelength region that exhibits \ce{NH3} features is dominated by unexplained lines. Therefore, other molecules that are not included in this analysis might explain the features as well.
    \item The isotopologues \ce{^{13}CO2} and \ce{^{13}CCH2} are clearly identified in the spectrum. The signatures for \ce{^{13}CH4} and \ce{^{13}CCH6} are not clearly visible and need further investigation.
    \item \ce{HCN} is emitting under higher temperatures (from $695^{+35}_{-40}\, \rm K$ to $820^{+70}_{-40}\, \rm K$) than all other detected molecules ($\lesssim 400\,\rm K$). We speculate that \ce{HCN} is emitting from a radially extended upper disk layer, while all other molecules are roughly following a common temperature power law of radial emission.
    \item The molecular emission conditions are well constrained for \ce{CO2}, \ce{C6H6}, \ce{C4H2}, \ce{C2H6}, and \ce{CH4}, with \ce{C2H2} being well constrained in temperature only and especially \ce{CH3} being poorly determined. In our retrieval, \ce{CO2} is emitting roughly co-spatially with other hydrocarbon molecules.
    \item A \ce{C4H2} quasi-continuum from about $15.5\,\rm \mu m$ to $16.4\,\rm \mu m$ is robustly identified.
    \item The optically thin dust of Sz\,28 consists mainly of large grains ($5\,\rm \mu m$) and shows a high crystallinity fraction ($\sim 45\%$) hinting to highly evolved dust in Sz\,28.
\end{itemize}

\begin{acknowledgements}
      We acknowledge funding from the European Union H2020-MSCA-ITN-2019 under grant agreement no. 860470 (CHAMELEON).
\end{acknowledgements}
\bibliographystyle{aa} 
\bibliography{lib.bib} 

\appendix

\section{Retrieval with restricted emitting areas\label{sec:restrict_radius}}

\begin{figure*}[p]
    \centering
    \includegraphics[width=0.97\linewidth]{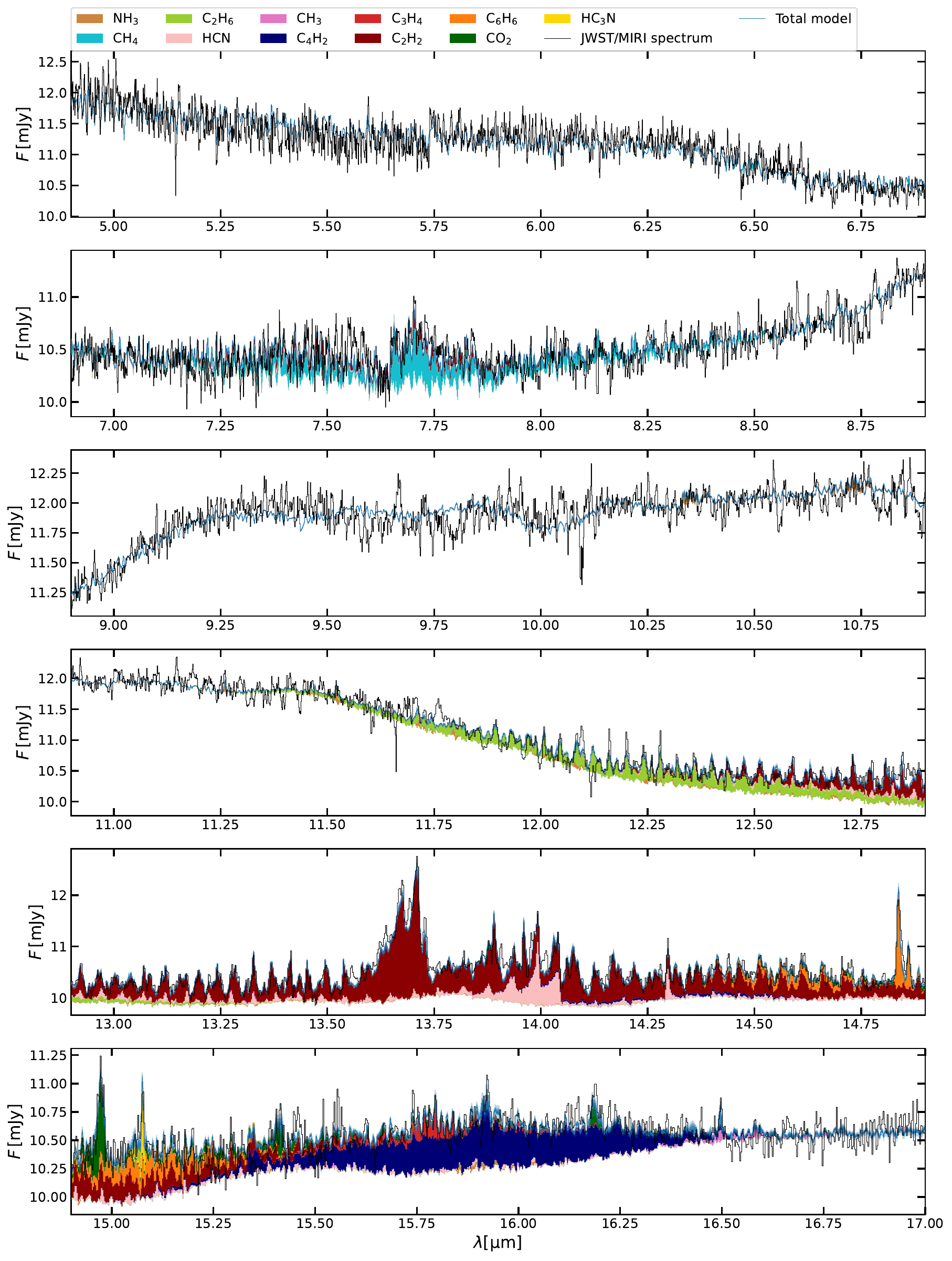}
    \caption{Molecular emission of the maximum likelihood model of Sz\,28 for the retrieval restricted to $r_{\rm eff}\leq2\,\rm au$. The JWST/MIRI spectrum is shown in black. The median flux of the posterior models is shown in blue with the $1\sigma$, $2\sigma$, and $3\sigma$ flux levels being displayed in lighter colours. The cumulative contributions of all included molecules are shown in different colours.}
    \label{fig:mol_spectrum_restricted}
\end{figure*}

The retrieval presented in Sect.~\ref{sec:final_fit} results in low temperatures from large emitting areas for \ce{NH3}, \ce{C2H6}, and \ce{C3H4}. 
To test the robustness of this solution, we reran the retrieval (with the same settings and priors) with an additional restriction on the emitting area. 
By setting the likelihood function to highly unlikely values for solutions with $r_{\rm eff}>2\,\rm au$, we excluded these cases from the retrieval.
The extracted parameters are listed in Table~\ref{tab:posterior_restrict}, with the fluxes from the posterior of models with their molecular contribution shown in Fig.\ref{fig:mol_spectrum_restricted}. As discussed in Section~\ref{sec:final_fit}, the restricted retrieval (Fig.~\ref{fig:mol_spectrum_restricted}) is able to describe the JWST/MIRI spectrum of Sz\,28 with a similar fitting quality as the unrestricted fit (Fig.~\ref{fig:mol_spectrum}). This shows that solutions with smaller emitting areas are possible.

\begin{table}[t]
    \caption{Posterior parameter values and uncertainties for selected parameters of the Sz\,28 fit for the retrieval restricted to $r_{\rm eff}\leq2\,\rm au$.}
    \centering
    \vspace{-2mm}
    \resizebox{!}{\linewidth}{
    \begin{tabular}{p{0.6cm} l|l|p{0.6cm}l|l}
\hline \hline & & & & & \\[-1.9ex] 
\multicolumn{2}{l|}{Parameter} & Posterior & \multicolumn{2}{l|}{Parameter} & Posterior \\ \hline  
  & & & & & \\[-1.9ex] 
\multicolumn{2}{c|}{$T_{\rm rim}$} & $977^{+11}_{-9}\, \rm K$ & &$r_{\rm eff}$ & $0.51^{+0.7}_{-0.29}\, \rm au$\\  
 & & & & & \\[-1.9ex] 
\multicolumn{2}{c|}{$T^{\rm sur}_{\rm min}$} & $281^{+20}_{-27}\, \rm K$ & &$t_{0.85}$ & $92^{+40}_{-24}\, \rm K$\\  
 & & & & & \\[-1.9ex] 
\multicolumn{2}{c|}{$T^{\rm sur}_{\rm max}$} & $620^{+500}_{-210}\, \rm K$ & $\rm C_3H_4$&$t_{0.15}$ & $144^{+50}_{-32}\, \rm K$\\  
 & & & & & \\[-1.9ex] 
\multicolumn{2}{c|}{$T^{\rm mid}_{\rm min}$} & $226^{+13}_{-11}\, \rm K$ & &$\Sigma_{0.85}$ & $8^{\times17}_{\div35}(+17)\, \rm cm^{-2}$\\  
 & & & & & \\[-1.9ex] 
\multicolumn{2}{c|}{$T^{\rm mid}_{\rm max}$} & $930^{+350}_{-340}\, \rm K$ & &$\Sigma_{0.15}$ & $1.4^{+7}_{-1.3}(+17)\, \rm cm^{-2}$\\  
 & & & & & \\[-1.9ex]\cline{4-5} 
 
\multicolumn{2}{c|}{$q_{\rm mid}$} & $-0.21^{+0.06}_{-0.05}$ & &$r_{\rm eff}$ & $0.056^{+0.007}_{-0.028}\, \rm au$\\  
 & & & & & \\[-1.9ex] 
\multicolumn{2}{c|}{$q_{\rm sur}$} & $-0.23^{+0.09}_{-0.16}$ & &$t_{0.85}$ & $278^{+15}_{-20}\, \rm K$\\  
 & & & & & \\[-1.9ex] 
\multicolumn{2}{c|}{$q_{\rm emis}$} & $-0.33^{+0.13}_{-0.09}$ & $\rm C_2H_2$&$t_{0.15}$ & $308^{+20}_{-20}\, \rm K$\\  
 & & & & & \\[-1.9ex] 
\multicolumn{2}{c|}{$a_{\rm obs}$} & $0.01437^{+0.00010}_{-0.00010}$ & &$\Sigma_{0.85}$ & $2.9^{\times1.9}_{\div318}(+20)\, \rm cm^{-2}$\\  
 & & & & & \\[-1.9ex] 
\cline{1-2} 
&$r_{\rm eff}$ & $0.25^{+0.8}_{-0.17}\, \rm au$ &&$\Sigma_{0.15}$ & $1.1^{\times5.1}_{\div122127}(+20)\, \rm cm^{-2}$\\  
 & & & & & \\[-1.9ex]\cline{4-5} 
 
&$t_{0.85}$ & $124^{+40}_{-28}\, \rm K$ &&$r_{\rm eff}$ & $0.18^{+0.08}_{-0.06}\, \rm au$\\  
 & & & & & \\[-1.9ex] 
$\rm NH_3$&$t_{0.15}$ & $210^{+120}_{-70}\, \rm K$ &&$t_{0.85}$ & $126^{+40}_{-17}\, \rm K$\\  
 & & & & & \\[-1.9ex] 
&$\Sigma_{0.85}$ & $6^{\times20}_{\div9.0}(+18)\, \rm cm^{-2}$ &$\rm C_6H_6$&$t_{0.15}$ & $239^{+27}_{-40}\, \rm K$\\  
 & & & & & \\[-1.9ex] 
&$\Sigma_{0.15}$ & $5^{+34}_{-4}(+17)\, \rm cm^{-2}$ &&$\Sigma_{0.85}$ & $3.6^{+1.7}_{-1.1}(+17)\, \rm cm^{-2}$\\  
 & & & & & \\[-1.9ex] 
\cline{1-2} 
&$r_{\rm eff}$ & $0.026^{+0.005}_{-0.004}\, \rm au$ &&$\Sigma_{0.15}$ & $2.6^{+0.9}_{-0.7}(+17)\, \rm cm^{-2}$\\  
 & & & & & \\[-1.9ex]\cline{4-5} 
 
&$t_{0.85}$ & $308^{+28}_{-26}\, \rm K$ &&$r_{\rm eff}$ & $0.064^{+0.034}_{-0.020}\, \rm au$\\  
 & & & & & \\[-1.9ex] 
$\rm CH_4$&$t_{0.15}$ & $398^{+50}_{-32}\, \rm K$ &&$t_{0.85}$ & $170^{+40}_{-40}\, \rm K$\\  
 & & & & & \\[-1.9ex] 
&$\Sigma_{0.85}$ & $4.2^{+4}_{-1.9}(+20)\, \rm cm^{-2}$ &$\rm CO_2$&$t_{0.15}$ & $280^{+80}_{-60}\, \rm K$\\  
 & & & & & \\[-1.9ex] 
&$\Sigma_{0.15}$ & $1.2^{+0.8}_{-0.5}(+20)\, \rm cm^{-2}$ &&$\Sigma_{0.85}$ & $1.8^{+11}_{-1.5}(+19)\, \rm cm^{-2}$\\  
 & & & & & \\[-1.9ex] 
\cline{1-2} 
&$r_{\rm eff}$ & $0.53^{+0.6}_{-0.23}\, \rm au$ &&$\Sigma_{0.15}$ & $5^{+20}_{-4}(+18)\, \rm cm^{-2}$\\  
 & & & & & \\[-1.9ex]\cline{4-5} 
 
&$t_{0.85}$ & $97^{+13}_{-14}\, \rm K$ &&$r_{\rm eff}$ & $0.27^{+0.31}_{-0.15}\, \rm au$\\  
 & & & & & \\[-1.9ex] 
$\rm C_2H_6$&$t_{0.15}$ & $125^{+22}_{-13}\, \rm K$ &&$t_{0.85}$ & $160^{+50}_{-50}\, \rm K$\\  
 & & & & & \\[-1.9ex] 
&$\Sigma_{0.85}$ & $6^{+26}_{-5}(+22)\, \rm cm^{-2}$ &$\rm HC_3N$&$t_{0.15}$ & $270^{+80}_{-60}\, \rm K$\\  
 & & & & & \\[-1.9ex] 
&$\Sigma_{0.15}$ & $3.3^{\times5.1}_{\div2031}(+20)\, \rm cm^{-2}$ &&$\Sigma_{0.85}$ & $2.8^{+16}_{-2.3}(+15)\, \rm cm^{-2}$\\  
 & & & & & \\[-1.9ex] 
\cline{1-2} 
&$r_{\rm eff}$ & $0.20^{+0.21}_{-0.11}\, \rm au$ &&$\Sigma_{0.15}$ & $3.8^{+12}_{-3.0}(+15)\, \rm cm^{-2}$\\  
 & & & & & \\[-1.9ex]\cline{4-5} 
 
&$t_{0.85}$ & $670^{+50}_{-50}\, \rm K$ &$\rm NH_3$&$\mathcal{N}_{\rm tot}$ & $1.8^{+240}_{-1.7}(+44)$\\  
 & & & & & \\[-1.9ex]\cline{4-5} 
 
$\rm HCN$&$t_{0.15}$ & $880^{+140}_{-80}\, \rm K$ &$\rm CH_4$&$\mathcal{N}_{\rm tot}$ & $1.4^{+1.4}_{-0.7}(+44)$\\  
 & & & & & \\[-1.9ex]\cline{4-5} 
 
&$\Sigma_{0.85}$ & $1.3^{+6}_{-1.0}(+15)\, \rm cm^{-2}$ &$\rm C_2H_6$&$\mathcal{N}_{\rm tot}$ & $3.2^{+26}_{-2.9}(+48)$\\  
 & & & & & \\[-1.9ex]\cline{4-5} 
 
&$\Sigma_{0.15}$ & $2.2^{+8}_{-1.7}(+15)\, \rm cm^{-2}$ &$\rm HCN$&$\mathcal{N}_{\rm tot}$ & $4.9^{+0.4}_{-0.5}(+40)$\\  
 & & & & & \\[-1.9ex]\cline{4-5} 
 
\cline{1-2} 
&$r_{\rm eff}$ & $0.024^{+0.05}_{-0.012}\, \rm au$ &$\rm CH_3$&$\mathcal{N}_{\rm tot}$ & $5^{\times64}_{\div34}(+43)$\\  
 & & & & & \\[-1.9ex]\cline{4-5} 
 
&$t_{0.85}$ & $230^{+110}_{-90}\, \rm K$ &$\rm C_4H_2$&$\mathcal{N}_{\rm tot}$ & $5^{+100}_{-4}(+44)$\\  
 & & & & & \\[-1.9ex]\cline{4-5} 
 
$\rm CH_3$&$t_{0.15}$ & $340^{+100}_{-110}\, \rm K$ &$\rm C_3H_4$&$\mathcal{N}_{\rm tot}$ & $4^{\times84}_{\div17}(+43)$\\  
 & & & & & \\[-1.9ex]\cline{4-5} 
 
&$\Sigma_{0.85}$ & $9^{+11000}_{-8}(+18)\, \rm cm^{-2}$ &$\rm C_2H_2$&$\mathcal{N}_{\rm tot}$ & $1.2^{+0.6}_{-0.5}(+44)$\\  
 & & & & & \\[-1.9ex]\cline{4-5} 
 
&$\Sigma_{0.15}$ & $2.5^{\times58}_{\div90}(+20)\, \rm cm^{-2}$ &$\rm C_6H_6$&$\mathcal{N}_{\rm tot}$ & $8^{+10}_{-5}(+42)$\\  
 & & & & & \\[-1.9ex]\cline{4-5} 
 
\cline{1-2} 
&$r_{\rm eff}$ & $0.092^{+0.06}_{-0.031}\, \rm au$ &$\rm CO_2$&$\mathcal{N}_{\rm tot}$ & $3.6^{+32}_{-2.8}(+43)$\\  
 & & & & & \\[-1.9ex]\cline{4-5} 
 
&$t_{0.85}$ & $129^{+25}_{-27}\, \rm K$ &$\rm HC_3N$&$\mathcal{N}_{\rm tot}$ & $1.1^{+4}_{-0.7}(+41)$\\  
 & & & & & \\[-1.9ex] 
$\rm C_4H_2$&$t_{0.15}$ & $174^{+21}_{-20}\, \rm K$ & & \\ & & & & & \\[-1.9ex] 
&$\Sigma_{0.85}$ & $2.3^{+22}_{-1.9}(+20)\, \rm cm^{-2}$ & & \\ & & & & & \\[-1.9ex] 
&$\Sigma_{0.15}$ & $1.0^{+1.2}_{-0.6}(+19)\, \rm cm^{-2}$ & & \\
\end{tabular}}
\tablefoot{\\
The notation $a(+b)$ means $a\times10^b$. For parameters that have a lower limit of more than an order of magnitude below the retrieved value, the uncertainties are provided as factors. $\mathcal{N}_{\rm tot}$ denotes the total number of molecules that contributed to the described emission and not the total number of molecules in the disk.}
    \label{tab:posterior_restrict}
\end{table}

\end{document}